\pdfoutput=1
\documentclass{article}

\usepackage[preprint]{neurips_2026}

\usepackage[utf8]{inputenc}
\usepackage[T1]{fontenc}
\usepackage{amsmath,amssymb}
\usepackage{booktabs,multirow}
\usepackage{graphicx}
\usepackage{enumitem}
\usepackage{microtype}
\usepackage{url}

\title{Right-Sizing LLM-Agent Decomposition in VAT Determination:\\
       A Pilot Controlled Sweep}

\author{%
  Pedro Santos\\
  Independent Researcher\\
  \texttt{pedromiguelbsantos@gmail.com}\\
  ORCID: 0009-0007-3191-9617
}

\begin{document}
\maketitle

\begin{abstract}
Recent LLM-agent systems make conflicting design bets: decompose work across many narrow agents, or use one strong tool-using agent. This pilot studies that choice on bounded cross-border VAT determination with reverse charge, where every case has an oracle label under a fixed rule set and the intermediate decisions --- classification, jurisdiction, rate, exemption, and reverse-charge logic --- are independently scoreable.

We hold the activity surface fixed --- subtasks, tools, I/O schemas, validation checks, orchestrator, base model, and merge policy --- and vary only the assignment of subtasks to workers across four orchestrated configurations, from one wide worker to five narrow ones, compared against S0, a tuned no-orchestrator ReAct/function-calling single agent, with a deterministic rule engine as oracle. The program comprises 4,400 runs: a 40-case, five-repeat main sweep, matched-token arms separating prompt-budget from agent-count effects, and three failure-injection arms, all evaluated against pre-registered falsification criteria.

The two intermediate configurations lead on accuracy (0.830, against endpoints at 0.720 and 0.770) but do not meet the pre-stated bar against the fine endpoint, so the intermediate-optimum hypothesis remains unsupported at pilot scale --- a directional result. The single agent does not Pareto-dominate the orchestrated set. The pre-registered matched-token criterion fires: the budget-matched single agent lands 6.5 points below the leading configuration, but the interval includes zero, so any advantage is reported as consistent with a prompt-budget explanation. Under injection, availability faults --- timeout and shared-tool outage --- are absorbed at every granularity, with wide-scope restart over-recovering its own baseline by +0.160, while a single schema-conforming hallucinated record degrades every configuration and inverts the ordering, hitting fragmented configurations hardest. The contribution is a bounded, preregistered pilot heuristic for right-sizing agent decomposition in VAT-style determination --- place one partition boundary at the dependency-layer midpoint --- released with oracle, dataset, harness, raw traces, and analysis pipeline.

\end{abstract}

\section{Introduction}

Recent LLM-agent systems make conflicting design bets about whether complex work should be decomposed across narrow agents or handled by one strong tool-using agent. Anthropic reports that an orchestrator--subagent research system improved an internal research benchmark by 90.2\% over a single Claude Opus 4 agent, at roughly 15$\times$ the token cost \citep{anthropic2025}. Cognition argues that multi-agent designs are fragile because context and implicit decisions are dispersed across agents \citep{cognition2025}. Recent single-agent work, including OneFlow and Operand Quant, further challenges the assumption that orchestration is necessary for strong performance \citep{xu2026oneflow,sahney2025operand}. At the same time, \citet{chen2024llmcalls} show that compound LLM systems can scale non-monotonically with the number of calls: more calls can help, then hurt.

The unresolved question is empirical: what changes when only agent scope changes? Many prominent comparisons vary the orchestration framework, base model, prompt budget, tool surface, and task at the same time. This leaves the narrower effect of agent-scope granularity under-controlled.

This paper reports a pilot controlled sweep of agent-scope granularity on cross-border VAT determination with reverse charge. VAT determination is a deliberate testbed choice. Each case has an oracle label under a bounded VAT rule set, making accuracy unambiguous within the harness. The workflow decomposes into auditable intermediate decisions --- classification, jurisdiction, rate, exemption, and reverse-charge logic --- each independently scoreable. Token cost and median latency are operationally meaningful metrics. The task combines parallelizable lookups and classifications with sequential dependency in the final reverse-charge decision.

We hold the \emph{activity surface} fixed: the required subtasks, tools, I/O schemas, validation checks, base model family, and merge policy remain constant. Across four orchestrated configurations C1--C4, we vary only how that surface is assigned to workers, from one wide orchestrated worker to increasingly narrow workers, under a preregistered merge policy. We compare these against \textbf{S0}, a tuned ReAct/function-calling single agent with no orchestrator. \textbf{C1 is not treated as a baseline}; it is the one-worker orchestrated control that separates orchestration overhead from no-orchestrator single-agent behavior --- a distinction usually conflated in the literature. A deterministic rule engine serves as oracle and validator, not as a competitor.

To address the concern that multi-agent gains may be prompt-budget gains, we run a matched-token comparison between S0 and a pre-specified orchestrated reference --- the dependency-layer midpoint --- with a post-selection sensitivity arm at the best-performing configuration. To probe robustness, we inject two isolated failure modes --- timeout and hallucinated output --- plus one shared-tool-outage condition that tests the boundary of localized fallback under correlated dependency failure.

\textbf{Findings.} The program comprises 4,400 valid runs evaluated against three pre-registered falsification criteria (\S6.6). The two intermediate configurations lead the main sweep at 0.830 accuracy against endpoints of 0.720 (wide) and 0.770 (fine), and error localization shows why: wide scopes fail earliest at jurisdiction determination, the finest scope at reverse-charge synthesis, with the intermediates minimizing the total. The pre-stated material bar against the fine endpoint is nonetheless not met, so the intermediate-optimum hypothesis remains unsupported for this workload --- a directional pilot reading, not evidence of absence. S0 does not Pareto-dominate the orchestrated set. The confirmatory matched-token contrast is $-0.065$ with a confidence interval including zero, so the pre-registered criterion fires and any C2 advantage is reported as consistent with a prompt-budget explanation, even though the point estimate favors C2. Under injection, availability faults are absorbed at every granularity --- with wide-scope restart over-recovering its own un-injected baseline by +0.160 --- while a single schema-conforming hallucinated record degrades every configuration, hits the fragmented ones hardest, and inverts the main-sweep ordering.

\textbf{Contributions.} (1) A controlled pilot methodology that holds the activity surface, tools, model family, I/O schema, orchestrator, and merge policy fixed across C1--C4, with S0 evaluated separately, three falsification criteria fixed before results inspection, and a fully released artifact --- oracle, dataset, harness, raw traces, and analysis pipeline. (2) Main-sweep results on 40 oracle-labeled cases with five stochastic repeats per condition: a non-monotonic granularity curve with both intermediates at 0.830, case-clustered bootstrap confidence intervals, effect sizes, and earliest-failing-subtask error localization. (3) A matched-token comparison whose pre-registered criterion fires, disciplining the prompt-budget-versus-agent-count reading. (4) A failure-injection taxonomy: availability perturbations --- timeout and shared-tool outage --- absorbed at every granularity, with a wide-scope restart overshoot, versus a content perturbation --- a schema-conforming hallucinated record --- that degrades all configurations and inverts the ordering.

\textbf{Research questions.} RQ1: how does orchestrated granularity affect accuracy, median latency, token cost, and error type? RQ2: which configurations are Pareto-efficient or dominated? RQ3: how much of the observed effect remains under a matched-token comparison with S0? RQ4: how does granularity affect localized fallback under isolated failures and shared-tool outage?

\textbf{Scope.} This is a pilot. We study one VAT-style workload, one base model family, four orchestrated granularity levels, and synthetic failure injections. The result is a bounded candidate heuristic for VAT-style determination workflows, not a general design rule for enterprise agents.

\section{Related Work}

We organize related work around four questions needed for a controlled agent-granularity sweep: when multi-agent orchestration helps, how prior systems choose role granularity, how compound LLM systems scale with calls and prompt budget, and how legal-critical agent systems handle validation and failure.

\textbf{Agent-proliferation debate.} Recent LLM-agent systems provide empirical pressure on both sides of the multi-agent debate. Magentic-One \citep{fourney2024magentic} reports a generalist orchestrator--worker architecture in which a lead orchestrator plans, tracks progress, and re-plans while delegating to specialized workers, with statistically competitive results on GAIA, AssistantBench, and WebArena; it serves as the closest architectural antecedent for our orchestrator. \citet{tian2025beyond} benchmark multi-turn multi-agent orchestration against strong single-LLM baselines on GPQA-Diamond, IFEval, and MuSR, including ablations on authorship visibility and vote observation. \citet{li2026skills} studies when multi-agent systems can be compiled into single-agent skill libraries, reporting a capacity-like phase transition in skill selection that hierarchical routing can mitigate. Magentic Marketplace \citep{bansal2025marketplace} extends this orchestration line to environment-level evaluation, contributing an open-source simulated two-sided market in which assistant agents representing consumers transact with competing service agents at scale. These studies motivate agent-count and role-scope questions, but they do not jointly fix the activity surface, orchestrator, tool set, schema, model family, and merge policy while varying only agent scope.

\textbf{Decomposition and role granularity.} Classical MAS treats decomposition as a design problem: role-based methodologies \citep{pavon2003ingenias,cossentino2005passi} and learned role-discovery systems show that the choice of role scope affects coordination and performance. RODE \citep{wang2020rode} decomposes joint action spaces into restricted role action spaces; Dynamic Role Discovery \citep{xia2023role} reports 55--60\% average win-rate drops after ablating role-discovery components, evidence that role granularity is causally important rather than incidental. \citet{karimadini2009decomposition} formalize the stronger point that local decomposition must preserve global behavior, so granularity choices are not arbitrary. In the LLM era, MetaGPT \citep{hong2023metagpt}, ChatDev \citep{qian2023chatdev}, and AutoGen \citep{wu2023autogen} popularized fixed narrow-role systems, while \citet{miyazaki2026trading} vary decomposition granularity in a financial trading domain and TDAG \citep{wang2025tdag} generates subagents dynamically while emphasizing error propagation. Microservices granularity studies \citep{verarivera2021granularity,mendonca2024granularity,zhao2025granularity} provide a methodological analogue: they treat unit scope as an independent variable and measure performance, cost, and energy under controlled decompositions. We borrow this controlled-experiment logic while treating probabilistic outputs, context-window coupling, and non-deterministic fallback as LLM-MAS-specific phenomena.

\textbf{Compound-system scaling and prompt-budget confounds.} \citet{chen2024llmcalls} derive scaling laws for compound inference systems and show that Vote and Filter-Vote can scale non-monotonically with call count: additional calls can improve performance and then degrade it. This motivates testing for a bounded optimum rather than assuming more agents are better. A separate concern is that multi-agent gains may reflect total prompt budget rather than agent count. OneFlow \citep{xu2026oneflow} argues that many homogeneous MAS workflows use the same base LLM and differ mainly by prompts, tools, and workflow position, and that strong single-agent workflows can match them at lower cost. Operand Quant \citep{sahney2025operand} adds a complementary single-agent challenge in ML-engineering settings, reinforcing that S0 must be a tuned tool-using baseline rather than a weak no-agent strawman. \citet{su2025difficulty} further suggest that workflow complexity should be matched to query difficulty rather than fixed globally. RQ3 addresses this confound through a matched-token comparison between S0 and the best orchestrated configuration.

\textbf{Tax/legal-critical agents and fault handling.} \citet{gogani2025synedrion} present Synedrion, the closest tax-prep agent system: a role-based LLM framework with metamorphic testing for legal-critical tax software, reporting that GPT-4o-mini inside the agentic system outperforms GPT-4o and Claude 3.5 on complex tax-code tasks (45\% worst-case pass rate vs.\ 9--15\%). Synedrion evaluates one architecture; we instead sweep granularity on a bounded VAT workflow with oracle labels. LegalBench \citep{guha2023legalbench} and SARA \citep{holzenberger2020sara} provide adjacent statutory-reasoning context, while rule-based tax engines motivate deterministic validation rather than learned baselines. On recovery, work on fault-tolerant ``graceful degradation'' \citep{lin2019degradation} distinguishes full recovery to the original specification from recovery to states satisfying only a critical subset of requirements; this pilot reports the narrower quantity of localized fallback under isolated injections. Large-scale failure taxonomies of LLM multi-agent systems \citep{cemri2025mast} further document coordination and recovery failures in deployed frameworks, but do not provide controlled measurements of fallback quality as a function of agent scope. RQ4 addresses that gap with timeout, hallucinated-output, and shared-tool-outage injections.

\textbf{Position.} To our knowledge, this is the first controlled pilot on VAT-style tax determination to vary only agent scope under a fixed activity surface --- holding the orchestrator, base model family, tools, I/O schema, and merge policy fixed --- while including a tuned no-orchestrator single-agent baseline (S0) kept distinct from the one-worker orchestrated condition (C1), a matched-token control, and a localized-fallback protocol.

\section{VAT Determination as a Testbed}

We use cross-border VAT determination with reverse charge as the testbed for the granularity sweep. The choice is methodological rather than domain-specific: VAT determination is bounded enough to support oracle-labeled evaluation, yet structured enough to expose the coordination trade-offs that motivate agent decomposition. In particular, it has three properties useful for this study. First, each case has an unambiguous label under the bounded rule set used in the harness, so accuracy is well-defined. Second, the task decomposes into a fixed sequence of intermediate decisions --- classification, jurisdiction, rate, exemption, and reverse-charge logic --- that can be scored independently. Third, token cost and median latency are meaningful metrics because tax determination is commonly executed at transactional volume and must preserve an auditable decision trace.

\subsection{Bounded VAT task}

Each case represents a structured cross-border transaction with invoice-level attributes, party attributes, and one or more line items. The input schema includes supplier country, customer country, customer VAT-registration status, B2B/B2C status, line-item descriptions, line-item amounts, and a bounded product/service classification vocabulary. The output is a structured VAT determination containing the applicable jurisdiction, rate treatment, exemption status, reverse-charge status, liable party, VAT amount or non-charging reason, and rule references.

The rule set is modeled on EU-style VAT treatment for domestic supply, intra-community B2B supply with reverse charge, B2C supply with rate differences, and mixed goods/services invoices. For reverse-charge scenarios, the harness includes rules modeled on general B2B place-of-supply and reverse-charge logic of the kind associated with Council Directive 2006/112/EC. The implementation is deliberately bounded: it does not attempt to encode the full VAT law of any Member State, special-sector schemes, or real-time legislative updates. Its purpose is to provide a controlled activity surface with oracle labels, not to serve as a deployable VAT engine, and the harness makes no legal-compliance claims.

\subsection{Activity surface}

We define the activity surface as the tuple $\mathcal{A} = (T, \mathcal{I}, \mathcal{O}, \mathcal{V}, \mathcal{F}, \mathcal{D}, \mathcal{R})$, where $T$ is the set of required subtasks, $\mathcal{I}$ and $\mathcal{O}$ are typed input and output schemas, $\mathcal{V}$ is the validation-check set, $\mathcal{F}$ is the tool set, $\mathcal{D}$ is the dependency partial order, and $\mathcal{R}$ is the fixed set of read-only reference materials available to the harness.

The required subtasks $T$ are fixed across all experimental conditions:

\begin{enumerate}[leftmargin=*,itemsep=1pt,topsep=2pt]
\item \textbf{Goods/services classification.} Classify each line item into the bounded product/service vocabulary used by the rule engine.
\item \textbf{Jurisdiction determination.} Determine the place of supply and applicable VAT jurisdiction using supplier country, customer country, customer VAT-registration status, B2B/B2C status, and the classification result.
\item \textbf{Rate lookup.} Retrieve the applicable rate treatment for the jurisdiction and classification.
\item \textbf{Exemption check.} Evaluate the line item against the bounded exemption table.
\item \textbf{Reverse-charge decision.} Determine whether VAT is charged by the supplier or self-accounted for by the customer, using the prior decisions and the transaction status.
\end{enumerate}

The tool set $\mathcal{F}$ is also fixed: a classification reference, a VAT-registration validity check, a rate-table lookup, and a rule-citation retrieval tool. Each subtask emits a structured JSON record containing the decision, supporting fields, and rule reference.

In this pilot, $\mathcal{R}$ consists of the bounded exemption table. Any additional static reference material must be declared in the released configuration and held fixed across all conditions. Unlike tools in $\mathcal{F}$, references in $\mathcal{R}$ are not callable external operations; they are static, read-only materials that condition decisions but do not execute computation.

Validation checks $\mathcal{V}$ enforce schema conformance, required-field presence, rule-citation presence, and internal consistency between the emitted decision and the cited rule. For example, a reverse-charge output must be consistent with the jurisdiction decision, customer status, and B2B/B2C classification; a rate output must correspond to an entry in the bounded rate table; and an exemption output must cite an exemption-table rule when an exemption is asserted.

The dependency relation $\mathcal{D}$ reflects both parallel and sequential structure. Line-item classification can be performed directly from the initial line-item fields and does not depend on jurisdiction, rate, exemption, or reverse-charge outputs. The VAT-registration validity check can be called directly from the customer VAT-registration field and is consumed by jurisdiction determination. Jurisdiction determination depends on classification and customer status. Rate lookup and exemption checking depend on the jurisdiction and classification decisions and can proceed in parallel once those inputs are available. The final reverse-charge decision depends on all prior subtasks. This structure makes VAT determination suitable for testing agent granularity: some work can be decomposed across narrow workers, while final correctness still depends on coordinated synthesis.

All orchestrated configurations C1--C4 and the no-orchestrator baseline S0 implement the same activity surface. The required subtasks, tools, schemas, validation checks, reference materials, and dependency structure do not change across conditions. Within the orchestrated sweep C1--C4, only the assignment of subtasks to workers changes; S0 is evaluated separately as a no-orchestrator comparator.

\subsection{Ground truth and dataset}

The dataset contains 40 synthetic VAT cases generated under the bounded rule set. The deterministic rule engine emits both the final VAT determination and the labeled intermediate decisions for each subtask. The rule engine is used as the oracle and validator for evaluation; it is not treated as a competing agentic baseline.

Cases are stratified across four scenario families, with ten cases per family:

\begin{enumerate}[leftmargin=*,itemsep=1pt,topsep=2pt]
\item \textbf{Domestic supply.} Supplier and customer are in the same jurisdiction, with standard rate or exemption treatment.
\item \textbf{Intra-community B2B with reverse charge.} Supplier and VAT-registered customer are in different jurisdictions, exercising reverse-charge logic.
\item \textbf{B2C with rate differential.} Customer status changes the applicable place-of-supply and rate treatment.
\item \textbf{Mixed goods/services invoices.} A single case includes line items requiring different classification, rate, or exemption handling.
\end{enumerate}

A stratified subset of cases --- at least one per scenario family, with additional coverage of cases that exercise reverse-charge eligibility, exemption handling, B2B/B2C transitions, and mixed goods/services classification --- is manually spot-checked after generation. The spot check is not used to alter model outputs; it is used only to verify that the oracle labels generated by the rule engine conform to the bounded rule specification.

\subsection{Auditable intermediate evaluation}

Because each subtask emits a structured record with a rule citation, the testbed scores both final determinations and intermediate decisions. Final-answer accuracy measures whether the resulting VAT treatment matches the oracle label. Step-level accuracy measures whether each subtask output matches the corresponding oracle intermediate label. Citation completeness measures whether each emitted decision includes the required rule reference. Consistency checks measure whether the trace is internally coherent, even when the final answer is incorrect.

A final VAT error can arise from different sources: misclassification, incorrect jurisdiction selection, wrong rate lookup, missed exemption, or incorrect reverse-charge synthesis. Because every configuration emits the same structured trace, the testbed can localize where errors enter as agent scope changes. The granularity sweep can therefore measure not only whether a configuration is correct, but also how its error profile changes when work is assigned to wider or narrower agents.

\subsection{Scope of the testbed}

The testbed is synthetic and bounded by design. It does not cover the full VAT rule space of any jurisdiction, real invoice parsing, supplier master-data defects, or production posting and reporting semantics. It also does not make legal-compliance claims.

Within those limits, the harness is appropriate for a pilot controlled sweep. It is rule-faithful enough to make oracle labels meaningful, structured enough to support auditable step-level evaluation, and narrow enough that the activity surface can be fully specified and held constant across all configurations. The resulting evidence should therefore be read as a bounded heuristic for VAT-style determination workflows, not as a general claim about tax automation or enterprise agents.

\section{Configurations}

This section defines the experimental conditions used in the granularity sweep. All orchestrated configurations C1--C4 implement the activity surface $\mathcal{A}$ defined in \S3.2. The no-orchestrator baseline S0 implements the same activity surface without an orchestrator. Across conditions, the required subtasks, tools, reference materials, schemas, validation checks, dependency structure, base model family, decoding settings, timeout threshold, and logging protocol are held fixed. Retry opportunities are fixed within each condition's natural repair unit and reported explicitly because S0 and C1--C4 expose different repair granularities.

Within the orchestrated sweep C1--C4, the experimental variable is the assignment of subtasks to workers. S0 is not part of that sweep; it is a separate no-orchestrator comparator that differs from C1--C4 in two respects simultaneously --- it has no task ledger, no progress ledger, and no dependency-ordered dispatch, and its repair unit is the whole trace rather than the subtask. S0 is therefore analyzed alongside C1--C4 rather than as a point on the granularity curve, and is kept distinct from C1 throughout (\S4.5).

\subsection{Agent definition}

An \emph{agent} in this study is a tuple

$$a = (r,\; P_r,\; F_r,\; S_r,\; \pi_r),$$

where $r$ is the role, $P_r$ is the role-scoped prompt, $F_r \subseteq \mathcal{F}$ is the tool-permission subset, $S_r \subseteq \mathcal{I} \cup \mathcal{O} \cup \mathcal{R}$ is the subset of input fields, prior outputs, and read-only reference materials visible to the agent, and $\pi_r$ is the interaction protocol governing how the agent receives inputs and returns structured outputs. Here $\mathcal{R}$ denotes the set of static reference materials available to the harness --- bounded lookup tables and reference documents that are read-only and not callable as tools. In this pilot, $\mathcal{R}$ consists of the bounded exemption table; any additional static reference material must be declared in the released configuration and held fixed across all conditions. $\mathcal{I}$ and $\mathcal{O}$ are as defined in \S3.2.

Two agents are treated as distinct when they differ in role, tool permissions, state scope, or interaction protocol. Prompt wording alone is not sufficient. This definition addresses the homogeneity critique of multi-agent systems \citep{xu2026oneflow}: in the narrower configurations, workers differ not only by prompt, but also by permitted tools, visible state, expected schema, and position in the dependency graph.

We distinguish \emph{tool permissions} $F_r$ from \emph{visible state} $S_r$. Tools are callable external operations from the fixed set $\mathcal{F}$ defined in \S3.2. Visible state is the subset of case fields, prior subtask outputs, and static reference materials that the agent receives as input. The two are kept conceptually separate throughout \S4.3 to avoid the impression that workers in narrower configurations are differently advantaged on tooling when they merely receive different state slices.

\subsection{Orchestrator and fixed controls}

All orchestrated configurations use the same Magentic-One-style orchestrator \citep{fourney2024magentic}. The orchestrator maintains two ledgers. The \emph{task ledger} records open subtasks for the current case and their dependency status. The \emph{progress ledger} records completed subtasks, emitted structured records, validation outcomes, retry counts, and unresolved failures.

For each case, the orchestrator dispatches subtasks according to the dependency partial order $\mathcal{D}$. Independent subtasks may be dispatched in parallel when $\mathcal{D}$ permits. If a worker output fails validation, times out, or emits an incomplete structured record, the orchestrator retries that subtask up to the fixed per-subtask retry budget. After the budget is exhausted, the case is marked with the corresponding failure status and the available trace is retained for error analysis. The orchestrator is implemented as deterministic control logic over these ledgers: because the dependency order $\mathcal{D}$ is fixed and known, dispatch, validation, and retry require no orchestrator-level model calls, and all model calls in C1--C4 are worker calls.

The following controls are invariant across C1--C4:

\begin{itemize}[leftmargin=*,itemsep=1pt,topsep=2pt]
\item the orchestrator implementation and ledger semantics;
\item the base model family, pinned model version, and decoding settings;
\item the tool set $\mathcal{F}$ and tool implementations;
\item the reference set $\mathcal{R}$;
\item the input and output schemas $\mathcal{I}, \mathcal{O}$;
\item the validation checks $\mathcal{V}$;
\item the dependency order $\mathcal{D}$;
\item the timeout threshold, the per-subtask retry budget, and the concurrency cap;
\item trace logging, token accounting, and latency measurement.
\end{itemize}

The orchestrator is therefore not a variable in the sweep. Only the partition of subtasks across workers changes.

\subsection{Orchestrated configurations C1--C4}

The five subtasks in $T$ are abbreviated as follows:

\begin{itemize}[leftmargin=*,itemsep=1pt,topsep=2pt]
\item \textbf{CLS:} goods/services classification;
\item \textbf{JUR:} jurisdiction determination;
\item \textbf{RAT:} rate lookup;
\item \textbf{EXM:} exemption check;
\item \textbf{RCH:} reverse-charge synthesis.
\end{itemize}

The four orchestrated configurations are defined by the following partitions of $T$:

\begin{center}
\begin{tabular}{@{}lcl@{}}
\toprule
Configuration & Worker count & Worker assignment \\
\midrule
\textbf{C1} & 1 & $\{\text{CLS, JUR, RAT, EXM, RCH}\}$ \\
\textbf{C2} & 2 & $\{\text{CLS, JUR}\}$, $\{\text{RAT, EXM, RCH}\}$ \\
\textbf{C3} & 3 & $\{\text{CLS}\}$, $\{\text{JUR}\}$, $\{\text{RAT, EXM, RCH}\}$ \\
\textbf{C4} & 5 & $\{\text{CLS}\}$, $\{\text{JUR}\}$, $\{\text{RAT}\}$, $\{\text{EXM}\}$, $\{\text{RCH}\}$ \\
\bottomrule
\end{tabular}
\end{center}

\begin{figure}[t]
\centering
\includegraphics[width=\linewidth]{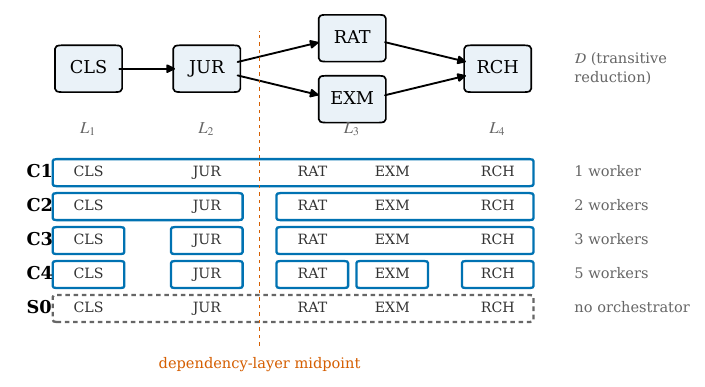}
\caption{The fixed activity surface and the experimental conditions. Top: the five subtasks and the transitive reduction of the dependency order $\mathcal{D}$, with dependency layers $L_1$--$L_4$. Bottom: worker partitions C1--C4 and the no-orchestrator baseline S0; the dashed vertical line marks the dependency-layer midpoint used by the merge policy (\S4.4) and the matched-token reference (\S6.3).}
\label{fig:surface}
\end{figure}

Figure~\ref{fig:surface} depicts the dependency order $\mathcal{D}$ and these partitions alongside S0.

For C1--C3, each worker receives the union of visible state and tool permissions required by the subtasks assigned to that worker; C4 is listed explicitly below because it is the finest decomposition and exposes the most restrictive permission slices.

\textbf{C1} is the one-worker orchestrated control. A single worker receives the full case, has access to the full tool set $\mathcal{F}$, and emits all five structured subtask records under the orchestrator's ledger and validation loop.

\textbf{C2} separates pre-rate determination from downstream tax treatment. The first worker performs classification and jurisdiction determination. The second worker receives the validated outputs from the first worker and performs rate lookup, exemption checking, and reverse-charge synthesis.

\textbf{C3} separates classification from jurisdiction while keeping downstream rate, exemption, and reverse-charge reasoning co-resident. This configuration tests whether isolating the two earliest decision points reduces error propagation without fully fragmenting the downstream synthesis step.

\textbf{C4} is the finest decomposition. Each worker is responsible for exactly one subtask and receives only the visible state required for that subtask. Tool permissions are restricted to the subset of $\mathcal{F}$ relevant to each subtask:

\begin{itemize}[leftmargin=*,itemsep=1pt,topsep=2pt]
\item the classification worker receives the line-item descriptions as visible state and the classification reference tool;
\item the jurisdiction worker receives supplier country, customer country, B2B/B2C status, customer VAT-registration status, and the classification output as visible state, and the VAT-registration validity check and rule-citation retrieval tools;
\item the rate worker receives the jurisdiction and classification outputs as visible state and the rate-table lookup tool;
\item the exemption worker receives the jurisdiction and classification outputs as visible state, the bounded exemption table from $\mathcal{R}$ as visible state, and the rule-citation retrieval tool;
\item the reverse-charge worker receives the case line items and all prior structured records as visible state, and the rule-citation retrieval tool. The line items are included because reverse-charge synthesis computes the VAT amount from the applicable rate and the line-item amounts; every field required by a subtask's contractual output is part of that subtask's visible state.
\end{itemize}

The bounded exemption table is treated as visible state from the reference set $\mathcal{R}$ rather than as a tool because it is a static reference distributed to whichever worker handles EXM, not a callable operation in $\mathcal{F}$. The fixed tool set $\mathcal{F}$ in \S3.2 --- classification reference, VAT-registration validity check, rate-table lookup, and rule-citation retrieval --- is therefore unchanged across all conditions.

C1 is not treated as a single-agent baseline. It is an orchestrated one-worker condition. Its comparison with S0 isolates the cost and behavior of the orchestrator itself. Comparisons among C1--C4 isolate the effect of changing worker scope under a fixed orchestrator.

\subsection{Merge policy}

The C1--C4 partitions follow a single pre-specified merge policy: \textbf{dependency-layer clustering}. Starting from the finest decomposition in C4, adjacent subtasks are merged only when doing so preserves the topological order of $\mathcal{D}$. Parallel-eligible subtasks that share immediate dependencies are kept co-resident in coarser configurations and separated only in the finest configuration. The policy was committed to the project repository under version control before any results from the main sweep were inspected.

This policy yields the progression from C4 to C1 as follows. C4 assigns one worker per subtask. C3 merges the downstream rate, exemption, and reverse-charge subtasks because they all depend on validated jurisdiction and classification outputs. C2 further merges classification with jurisdiction, producing a pre-treatment worker and a downstream treatment worker. C1 merges all subtasks into one wide orchestrated worker.

Two alternative policies were considered before the sweep but excluded from the primary analysis: \textbf{tool-locality clustering}, which groups subtasks by shared tools, and \textbf{semantic-similarity clustering}, which groups subtasks by overlap in rule families. Including them would introduce a second experimental variable. The primary study therefore evaluates granularity under one fixed, dependency-respecting merge policy; robustness across merge policies is deferred to follow-on work and discussed as a limitation in \S10.

\subsection{No-orchestrator baseline S0}

S0 is a tuned ReAct/function-calling single agent with no orchestrator, no task ledger, and no progress ledger. It receives the full case, has access to the full tool set $\mathcal{F}$ and the full reference set $\mathcal{R}$, and emits the same structured trace as the orchestrated configurations. Its output is evaluated using the same schemas and validation checks.

S0 is designed as a strong comparator rather than a weak strawman. It uses:

\begin{enumerate}[leftmargin=*,itemsep=1pt,topsep=2pt]
\item the same base model family and decoding settings as C1--C4;
\item the same tools and tool implementations;
\item the same input and output schemas;
\item the same validation checks $\mathcal{V}$, applied after trace emission;
\item up to three whole-trace repair attempts after validation failure, which is the natural repair unit available without an orchestrator;
\item a prompt tuned on a held-out development split separate from the 40 evaluation cases.
\end{enumerate}

The held-out development split contains eight cases drawn from the same generator as the evaluation set. These cases are used only for prompt tuning and are excluded from all reported evaluation results. The final S0 prompt, tuning budget, and development-case identifiers are reported in the appendix.

\textbf{Fairness principle for retries.} Each condition is granted equal retry opportunities at its natural repair unit: subtask-level for C1--C4, whole-trace-level for S0. This creates different worst-case call counts across conditions. We therefore treat retry-induced call volume as part of the measured cost of each architecture, rather than forcing an artificial equal-call cap that would distort either S0 or the orchestrated configurations.Per-condition retry counts, tool calls, and token usage are recorded per run in the released artifact; \S7 reports total token cost as the primary cost metric.

S0 and C1 are kept separate throughout the analysis. S0 measures no-orchestrator single-agent behavior. C1 measures one-worker behavior under the orchestrator, ledgers, validation loop, and retry policy. Conflating these conditions would mix orchestration overhead with agent-scope effects.

\subsection{Execution controls and logging}

All conditions use fixed decoding settings: temperature $0.2$ and a fixed maximum-token cap per call. Nucleus sampling (top-$p$) is left at the provider default because the provider API rejects requests that set temperature and top-$p$ simultaneously; sampling is therefore controlled through temperature alone. Tool ordering, case ordering, and validation execution are deterministic. The provider API does not expose decoding-seed control, so stochastic variance is estimated from repeated executions under the same decoding settings.

Each (condition, case) pair is run five times to support bootstrap confidence intervals over stochastic and case-level variance (\S6). The timeout threshold is fixed at 120 s per model call across all conditions.

Concurrency is capped so that parallel-eligible workers in C2--C4 do not contend for provider-side rate-limit capacity. This avoids attributing provider queuing delays to agent granularity. Wall-clock latency is measured from case submission to final validated trace or terminal failure. Token usage is logged per call and aggregated per case. The logs also record tool calls, validation failures, retry counts, timeout events, emitted rule citations, and final error categories.

The main sweep uses one pinned base model family. Heterogeneous-model variants are not run in this pilot; their omission is treated as a scoping decision and discussed in \S10 (Threats to Validity).

\section{A Chain-Error Model}

We introduce a small analytical model to state prior expectations for the granularity sweep and to give \S7 a vocabulary for interpreting observed patterns. The model is \textbf{interpretive, not predictive}: we do not use it to forecast configuration-level results, and we do not fit it to held-out configurations before reporting the sweep. Following \citet{chen2024llmcalls}, we treat the relationship between decomposition and end-to-end quality as potentially non-monotonic rather than assuming that more calls or more workers necessarily improve performance.

\subsection{Chain-error baseline and cross-step dependence}

Each orchestrated configuration implements the activity surface $\mathcal{A}$ at fixed logical depth $d = 5$ corresponding to the required subtasks $T = \{\text{CLS}, \text{JUR}, \text{RAT}, \text{EXM}, \text{RCH}\}$. The worker count $k$ varies across C1--C4; the logical depth does not. Let $C_i \in \{0,1\}$ indicate whether subtask $i$ emits the oracle-correct structured decision under configuration $c$, with marginal correctness probability $p_i^{(c)} = P(C_i = 1 \mid c)$, and let $A^{(c)} = P(C_1 = 1, \ldots, C_d = 1 \mid c)$ denote trace-level case accuracy. Under independent per-step correctness and no retry repair, $A^{(c)}_{\mathrm{indep}} = \prod_{i=1}^{d} p_i^{(c)}$. Because all five subtasks are required, improvements to any required subtask affect trace-level accuracy multiplicatively.

The independence baseline is a reference point, not an assumption. Errors in LLM-agent pipelines are rarely independent: a wrong classification can bias jurisdiction, rate, exemption, and reverse-charge reasoning, while shared visible state and handoff dependencies can induce correlated failures across workers. Observed trace accuracy can therefore deviate from $\prod_i p_i^{(c)}$ in either direction --- upward when errors cluster on the same hard cases, downward when they are dispersed or introduced by handoff-specific failures. We do not estimate a single correlation parameter in this pilot. \S7 reports error-type breakdowns by earliest failing subtask; full step-error co-occurrence tables ship with the released analysis outputs.

S0 is not a point on the C1--C4 granularity curve. It implements the same activity surface without an orchestrator, ledgers, or subtask-level repair. We therefore use the model primarily for C1--C4 and treat the S0--C1 contrast separately in \S5.4.

\subsection{Worker scope}

Worker scope affects the marginal terms $p_i^{(c)}$, and the expected sign is not fixed. Wider workers have access to more neighboring context and may perform better on synthesis-heavy subtasks --- especially RCH, which depends on all prior decisions --- but longer prompts and broader visible state can also increase interference between unrelated rule contexts. Narrower workers reduce role scope, tool access, and visible state, which can improve local decision quality by limiting irrelevant context, but can also fragment information across handoffs and increase dependence on inter-worker message quality. Different subtasks may therefore prefer different scopes within the same configuration. The sweep is designed to measure these effects rather than impose a monotonic expectation.

\subsection{Retry repair}

Validation and retry change effective step accuracy for failures that are detectable and repairable within budget. Let $q_i^{(c)}$ denote the conditional probability that an initially incorrect or invalid subtask output for $i$ is detected by $\mathcal{V}$ and successfully repaired within the retry budget under configuration $c$. The effective post-retry step accuracy is

$$\tilde{p}_i^{(c)} = p_i^{(c)} + \bigl(1 - p_i^{(c)}\bigr)\,q_i^{(c)}.$$

This expression folds together detectability by validation and successful correction after retry; it therefore applies only to validation-visible failures. Silent semantic errors that satisfy the schema and cite a plausible rule may not be repaired even when they are wrong under the oracle.

For C1--C4, repair occurs at the subtask level exposed by the orchestrator's ledgers: a failed CLS record can be retried before JUR consumes it; a failed RAT record can be retried without regenerating the full trace. S0 differs structurally. It emits a complete trace and receives whole-trace repair attempts after validation failure, so its repair effect is better represented at the joint-trace level: $\tilde{A}^{(S0)} = A^{(S0)} + \bigl(1 - A^{(S0)}\bigr)\,Q^{(S0)}$. Critically, $Q^{(S0)}$ folds together detection, \emph{localization}, and correction in a way that subtask-level $q_i^{(c)}$ does not: orchestrated repair benefits from validation-driven localization of the failing step, while S0 retries the entire trace whether the original failure was in CLS or RCH. This asymmetry, not the per-call retry budget, is the substantive difference between S0 and C1 in the model.

\subsection{Cost, latency, and qualitative expectations}

The logical dependency order $\mathcal{D}$ is fixed across C1--C4, but the execution graph induced by the worker partition is not. Finer decompositions expose parallel fanout --- especially for RAT and EXM, which are parallel-eligible once jurisdiction and classification are available --- and add subtask-level validation points, but also multiply model calls, handoff messages, and orchestration overhead. Wider decompositions reduce handoff overhead but may serialize work inside a larger call and increase prompt and context size. Granularity should therefore trace out a non-trivial cost-quality response rather than collapse to a single dominant point; \S7 measures token cost and wall-clock latency directly and reports them per configuration.

Three expectations follow for interpreting the sweep.

First, accuracy may peak at an intermediate worker count. Very wide workers may benefit from shared context but suffer from prompt interference and coarse repair. Very narrow workers may improve local focus and subtask-level repair but introduce handoff errors and fragmented context. Intermediate decompositions can therefore dominate either extreme when the gains from local focus and repair exceed the costs of coordination.

Second, the cost-quality frontier may exclude some configurations. A configuration that adds calls, retries, or handoffs without a corresponding accuracy gain should be Pareto-dominated; conversely, a configuration with slightly lower accuracy may remain relevant if it substantially reduces token cost or latency.

Third, the S0--C1 contrast estimates the \emph{combined} effect of adding the orchestrator, ledgers, validation loop, and subtask-level repair at one-worker scope. It does not isolate ``the orchestrator'' as a single mechanism. Comparisons among C1--C4 estimate the effect of changing worker scope under a fixed orchestrator; the S0--C1 comparison estimates the effect of moving from no-orchestrator to orchestrated execution at fixed scope.

These expectations define the interpretive vocabulary used in \S7 and revisited in \S9. They are not claims about what the results will show.

\section{Methodology}

This section specifies metrics, the baseline and oracle, the matched-token control for RQ3, the failure-injection protocol for RQ4, statistical analysis, falsification criteria stated before results, and reproducibility artifacts. The dataset (40 cases, \S3.3), configurations (C1--C4 and S0, \S4), and execution controls (\S4.6) are defined earlier and are not repeated here.

\subsection{Metrics}

Three metric families are measured per (condition, case, repeat): accuracy, cost, and latency.

\textbf{Accuracy.} \emph{Final-answer accuracy} is the indicator that the emitted VAT determination matches the oracle label across the structured output fields enumerated in \S3.1: jurisdiction, rate treatment, exemption, reverse-charge status, liable party, and VAT amount or non-charging reason. \emph{Step-level accuracy} is the indicator that each subtask record matches its oracle intermediate label, scored per subtask in $T$: CLS, JUR, RAT, EXM, and RCH. \emph{Trace consistency} is the indicator that all validation checks in $\mathcal{V}$ pass on the emitted trace; rule-citation completeness and citation--decision consistency are scored as components of trace consistency, following \S3.4. \emph{Error type} attributes each incorrect trace to its earliest failing subtask, supporting the error-propagation analysis for RQ1. If multiple subtasks at the same dependency layer fail, all are recorded for error-type analysis; for single-label summary plots, we use the fixed order CLS, JUR, RAT, EXM, RCH.

\textbf{Terminal-failure scoring.} A case is \emph{terminally failed} under a condition if its trace cannot be brought to validation pass within the condition's natural repair budget: subtask-level for C1--C4 and whole-trace for S0. Terminal failures are scored as final-answer incorrect and trace-inconsistent. Step-level fields not emitted because of terminal failure are marked missing for step-level reporting and counted as incorrect in end-to-end trace accuracy. Token cost and latency for terminal failures are reported in full, including all retries up to the budget.

\textbf{Cost.} \emph{Token cost} is the sum of prompt and completion tokens across all model calls for the case, including retries. It is the primary cost metric because it captures the retry asymmetry between S0 and C1--C4 (\S4.5). \emph{Tool calls} and \emph{retry counts} are reported separately to support interpretation. \emph{Dollar cost} is computed from a single pinned price sheet to permit comparison; it is treated as derived from token cost rather than as an independent metric.

\textbf{Latency.} \emph{Wall-clock latency} is measured from case submission to final validated trace or terminal failure. We report median latency per condition with bootstrap confidence intervals. We deliberately do not report p95/p99 latency: with $40 \times 5 = 200$ observations per condition, tail-percentile estimates are too noisy for reliable inference. Tail behavior is deferred to follow-on work with larger trial counts.

\subsection{Baseline and oracle}

One baseline and one oracle anchor the comparisons. \textbf{S0} (\S4.5) is the tuned no-orchestrator ReAct/function-calling single-agent baseline; it is the comparator against which the combined effect of the orchestrator, ledgers, validation loop, and subtask-level repair is measured.

The \textbf{deterministic rule engine} (\S3.3) is the oracle. It generates ground-truth labels and intermediate evaluation targets. It is not treated as a competing agentic baseline, does not run under stochastic repeats, and does not appear in cost or latency comparisons. The oracle labels are used for evaluation and analysis; they are not exposed to the LLM agents during generation or repair.

\subsection{Matched-token comparison (RQ3)}

To separate worker-count effects from total-prompt-budget effects, we run a matched-token comparison between S0 and orchestrated configurations on the same 40 cases with five stochastic repeats. The procedure has two parts.

\textbf{Pre-specified comparison.} Before inspecting main-sweep results, we designate \textbf{C2}, the dependency-layer midpoint, as the orchestrated reference for the matched-token control. We measure mean total token cost per case for C2 on the main sweep, $\bar{B}_{C2}$, then re-tune S0 to a budget cap yielding per-case token usage within $\pm 10\%$ of $\bar{B}_{C2}$ on the held-out development split (\S4.5). Additional budget is allocated across role description, exemplars, and intermediate scratchpad. We denote this variant $\text{S0}'_{C2}$ and compare it against C2 under the inferential procedure of \S6.5. This is the confirmatory matched-token comparison.

\textbf{Post-selection sensitivity.} After the main sweep, we identify the orchestrated configuration with highest mean final-answer accuracy, $C^\star$, repeat the matched-token tuning to budget $\bar{B}_{C^\star}$, and denote the resulting variant $\text{S0}'_{C^\star}$. Because $C^\star$ is selected on the accuracy outcome it is then tested on, this comparison is reported as a post-selection sensitivity analysis, not as a confirmatory test.

For both matched-token variants, token matching is performed only on the development split. Evaluation-case token usage is reported as observed. If evaluation token usage falls outside the $\pm 10\%$ target band, we report the deviation and do not retune on the evaluation cases.

The matched-token analysis estimates how much of any S0--orchestrated accuracy gap remains once prompt budget is equalized. It does not isolate every confound. Residual differences, including orchestrator ledgers, subtask-level repair, and validation-driven failure localization, are discussed in \S9.

\subsection{Failure-injection protocol (RQ4)}

We inject three failure types into each condition and measure localized fallback. Injections are applied to a separate copy of the 40-case set so that injected and un-injected results are paired by case identity. For each case, a single \emph{target subtask identity} $\tau \in T$ is sampled deterministically from a fixed seed. The same $\tau$ is used across all conditions and repeats for that case, so injection sites are paired across conditions.

The injection unit is the \emph{worker invocation responsible for $\tau$}, not necessarily a single model call to a single subtask. Under C4, the worker covers exactly $\tau$. Under C2 and C3, the responsible worker may cover several subtasks including $\tau$, and the entire invocation is affected. Under C1, the single worker covers all subtasks and is therefore the affected invocation. For S0, the injection unit is the trace-generation call whose emitted trace would cover the subtask-$\tau$ slot. This convention sacrifices strict per-call comparability in order to respect the worker scope induced by each configuration.

\textbf{Timeout (isolated).} The worker invocation responsible for $\tau$ is forced to time out at the fixed timeout threshold (\S4.6). For S0, the whole trace-generation call that would emit the subtask-$\tau$ slot is forced to time out. The orchestrator and S0 then exercise their natural repair mechanisms within their per-condition retry budgets. In implementation, the forced timeout is raised immediately rather than by holding the call open to the fixed 120 s threshold: the affected worker's state is identical to that of a genuine threshold timeout --- the task bundle is delivered and no assistant turn completes --- while avoiding threshold-length idle waits across the arm. None of the reported injection metrics depends on wall-clock latency, so this substitution leaves them unchanged; wall-clock latency under timeout injection is consequently not reported.

\textbf{Hallucinated output (isolated).} Under C1--C4, the structured record for subtask $\tau$ is replaced, after worker emission and before validation, with a schema-conforming but oracle-incorrect record carrying a plausible-but-wrong rule citation. Under S0, the same substitution is applied: after the emitted trace is produced and before validation, the subtask-$\tau$ slot of the trace is overwritten with the same schema-conforming oracle-incorrect record. The injected record is constructed deterministically from the case so that all repeats receive the same injection.

This condition tests silent semantic-error propagation. When the injected record satisfies $\mathcal{V}$, no validation-triggered retry fires. Recovery is only possible if downstream subtasks or cross-subtask consistency checks expose a contradiction. Otherwise, the trace may validate while remaining oracle-incorrect, especially when $\tau$ is the terminal subtask RCH or when the condition lacks a downstream worker with the responsibility and visible state to expose the contradiction.

\textbf{Shared-tool outage (correlated).} Each scenario family of ten cases is split into two blocks of five, yielding eight blocks across the 40-case evaluation set. For one deterministically selected case ID per block, fixed by seed, the rate-table lookup tool returns a transient error on the first invocation reaching RAT and recovers on subsequent attempts. Thus $m = 1$ for the shared-tool-outage condition, with eight affected cases (20\% of the evaluation set). Selection of affected cases is independent of execution scheduling, so the outage is paired across all configurations and repeats.

This condition tests a correlated dependency failure rather than localized worker failure: every configuration that reaches RAT on an affected case observes the same unavailable external tool on the first RAT lookup attempt. Recovery therefore depends on retry timing and per-condition fallback policy rather than sibling-worker substitution.

For each (injection, configuration) cell, we report:

\begin{itemize}[leftmargin=*,itemsep=1pt,topsep=2pt]
\item \emph{substitution success rate}: the fraction of injected cases reaching a validated trace within the per-condition repair budget;
\item \emph{all-case accuracy under injection}: final-answer accuracy computed over all injected cases, with terminal failures and validated-but-wrong traces both counted as incorrect;
\item \emph{validated-trace accuracy under injection}: final-answer accuracy on the subset reaching a validated trace, reported as a secondary descriptive metric;
\item \emph{cost penalty}: additional tokens relative to the un-injected baseline on the paired case.
\end{itemize}

All-case accuracy is the primary degradation metric so that terminal failures are not separated from validated-but-wrong traces in a way that hides total degradation.

For the hallucinated-output condition we additionally report the \emph{record substitution rate}: the fraction of injected cases in which the injected record is subsequently replaced by a repair before the final trace. This quantity separates exposure of the poisoned record from mere trace validation, since an injected record that satisfies $\mathcal{V}$ can validate without ever being corrected; it is undefined for the timeout and shared-tool-outage conditions. We also report, for each (injection, configuration) cell, the paired per-case accuracy delta relative to its un-injected main-sweep baseline, with case-clustered bootstrap confidence intervals computed as in \S6.5. The record substitution rate and the paired deltas are descriptive quantities outside the pre-specified headline family of \S6.5.

\subsection{Statistical analysis}

Because the same 40 cases appear across all conditions with five stochastic repeats each, headline comparisons are paired by case and clustered within case.

\textbf{Primary procedure.} We first aggregate the five repeats per (condition, case) to case-level summaries: mean for accuracy, mean for token cost, and median for latency. For paired comparisons between conditions, we use a \textbf{case-clustered paired bootstrap} with 1,000 resamples drawn at the case level. Each resample computes the paired difference in the statistic of interest: mean accuracy, mean token cost, or median latency across case-level summaries. Reported intervals are 95\% percentile bootstrap CIs on the paired difference.

For named pairwise hypotheses, we additionally report \textbf{paired permutation tests} over case-level summaries, using 1,000 permutations of within-case condition labels, with \textbf{Holm--Bonferroni} correction across the family of headline tests.

\textbf{Effect sizes.} We report \textbf{Cohen's $d_z$} for headline paired comparisons. The sign of $d_z$ follows the direction of the named contrast. For superiority claims, the small-effect threshold is directional: a configuration is treated as materially better only when the paired bootstrap CI for the accuracy difference excludes zero in the positive direction and $d_z \geq 0.2$.

\textbf{Descriptive supplement.} Mann--Whitney U is computed as a robustness check; values ship with the released analysis tables. It is not used for confirmatory inference.

\textbf{Pre-specified family of headline tests.} The pre-specified family, fixed before results inspection, is:

\begin{enumerate}[leftmargin=*,itemsep=1pt,topsep=2pt]
\item S0 vs. C1: orchestrated-control contrast at one-worker scope;
\item C1 vs. C4: granularity endpoints under fixed orchestrator;
\item C2 vs. C1: dependency-layer midpoint vs. wide endpoint under fixed orchestrator;
\item $\text{S0}'_{C2}$ vs. C2: confirmatory matched-token comparison for RQ3.
\end{enumerate}

Holm--Bonferroni correction is applied across these four tests. All other comparisons, including best-among-C1--C4 versus worst-among-C1--C4 cost-quality contrasts and the post-selection sensitivity $\text{S0}'_{C^\star}$ vs. $C^\star$, are reported descriptively outside the corrected family.

We note explicitly that 40 cases with five repeats each is a pilot-scale sample. We do not run an a priori power analysis; we report paired CIs and effect sizes and let readers judge inferential weight. Wide CIs are reported as wide, not interpreted as null findings. Non-significant results in the headline family should be read as inconclusive rather than as evidence of no effect.

\subsection{Falsification criteria}

We pre-state three criteria, fixed before results inspection. For accuracy comparisons, \emph{materially outperforms} means that the paired bootstrap CI for the accuracy difference excludes zero in the positive direction and the paired effect size satisfies $d_z \geq 0.2$.

\begin{enumerate}[leftmargin=*,itemsep=1pt,topsep=2pt]
\item \textbf{Intermediate-optimum granularity (RQ1).} If neither intermediate configuration, C2 nor C3, materially outperforms both endpoints, C1 and C4, on final-answer accuracy, the intermediate-optimum version of the non-monotonic-granularity hypothesis is unsupported for this workload. If the C1--C4 accuracy sequence is monotonic in either direction, this criterion is automatically unmet.
\item \textbf{Orchestration benefit (RQ2).} If S0 weakly Pareto-dominates all of C1--C4 on the token-cost/final-answer-accuracy plane under main-sweep conditions, the orchestrated multi-worker hypothesis is unsupported for this workload.
\item \textbf{Prompt-budget confound (RQ3).} If the $\text{S0}'_{C2}$ vs. C2 accuracy difference is not significant under the paired permutation test with Holm--Bonferroni correction and its paired bootstrap CI includes zero, we report that any C2 main-sweep advantage is consistent with a prompt-budget explanation rather than an agent-count explanation. All three criteria, the harness, and the analysis plan were committed under version control before any results were inspected; the released artifact's commit history provides timestamped evidence.
\end{enumerate}

These criteria are not the only signals in the results, but they discipline how headline claims are reported in \S7.

\subsection{Reproducibility}

The released artifact contains: the orchestrator implementation; prompt sets for C1--C4, S0, $\text{S0}'_{C2}$, and $\text{S0}'_{C^\star}$; the deterministic rule oracle; the 40 evaluation cases and 8 development cases with oracle labels; the failure-injection seeds, target-subtask assignments, affected-case lists, and shared-tool-outage setting $m = 1$; pinned model identifiers and decoding settings; per-(condition, case, repeat) raw traces; and the analysis scripts that produce all reported tables and figures from the raw traces.

The harness follows an AutoGenBench-style controlled-isolation pattern: each (condition, case, repeat) executes in a clean process with no shared mutable state. The supplementary package is prepared in two forms: an anonymized version for double-blind venues and an attributed version for non-anonymous venues, with author-identifying metadata restored after review or at submission to a non-anonymous venue.

\section{Results}

We report 4,400 valid runs: the main sweep (five conditions $\times$ 40 cases $\times$ 5 repeats = 1,000), two matched-token arms (2 $\times$ 200), and three failure-injection arms (3,000; reported in \S8). All numbers are produced by the released analysis pipeline from the released raw traces (\S6.7); headline inference follows \S6.5, and the pre-stated falsification criteria of \S6.6 discipline every verdict. Three harness defects were discovered and repaired during execution; the affected arms were re-run under the corrected harness, superseded attempts are archived in the released artifact and are not reported as results, and the defects are disclosed in \S10.2.

\subsection{Main sweep (RQ1, RQ2)}

Table~\ref{tab:main} reports the main sweep. All intervals are 95\% case-clustered bootstrap CIs (\S6.5); latency is the median of case-level medians.

\begin{table}[t]
\centering
\caption{Main sweep: final-answer accuracy, token cost, and median latency per condition, with 95\% case-clustered bootstrap CIs (\S6.5). Latency is the median of case-level medians.}
\label{tab:main}
\footnotesize
\setlength{\tabcolsep}{3.5pt}
\begin{tabular}{@{}lccccc@{}}
\toprule
Condition & Accuracy [95\% CI] & Tokens [95\% CI] & Med.\ latency, s [95\% CI] & Trace cons. & Term.\ fail. \\
\midrule
S0 & 0.755 [0.665, 0.840] & 11{,}899 [11{,}217, 12{,}628] & 11.8 [11.0, 12.7] & 1.000 & 0/200 \\
C1 & 0.720 [0.625, 0.810] & 11{,}317 [10{,}599, 12{,}000] & 12.1 [11.3, 12.6] & 1.000 & 0/200 \\
C2 & 0.830 [0.750, 0.910] & 14{,}590 [12{,}971, 16{,}416] & 15.8 [15.2, 16.3] & 0.985 & 3/200 \\
C3 & 0.830 [0.745, 0.915] & 13{,}591 [12{,}507, 14{,}928] & 16.7 [16.2, 17.4] & 0.985 & 3/200 \\
C4 & 0.770 [0.685, 0.855] & 15{,}845 [15{,}457, 16{,}238] & 18.4 [17.8, 19.7] & 1.000 & 0/200 \\
\bottomrule
\end{tabular}
\end{table}
\begin{figure}[t]
\centering
\includegraphics[width=0.9\linewidth]{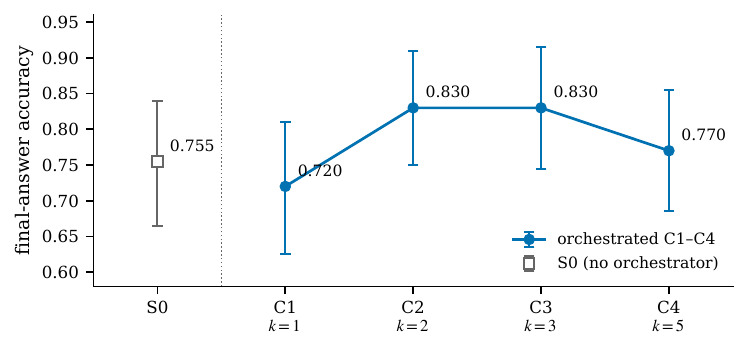}
\caption{Main-sweep final-answer accuracy by condition with 95\% case-clustered bootstrap CIs (Table~\ref{tab:main}). The orchestrated sequence C1--C4 is non-monotonic; S0 is plotted separately because it is not a point on the granularity curve (\S4.5). $n = 40$ cases $\times$ five repeats (200 runs per condition).}
\label{fig:curve}
\end{figure}

The accuracy sequence over C1--C4 is non-monotonic: 0.720, 0.830, 0.830, 0.770 (Figure~\ref{fig:curve}). The two intermediate configurations tie at the top, consistent with their sharing the identical downstream $\{$RAT, EXM, RCH$\}$ worker (\S4.3); separating JUR from CLS --- the only difference between C2 and C3 --- leaves accuracy unchanged. Median latency rises monotonically with worker count, from $\approx$12 s (S0, C1) to 18.4 s (C4), reflecting sequential dispatch along the dependency order. C1 is the cheapest condition in tokens and C4 the most expensive; dollar cost is derived from the pinned price sheet and tracks tokens (\S6.1). Terminal failures occur only in the intermediate configurations (three runs each, 1.5\%, all validation-budget exhaustion).

Table~\ref{tab:headline} reports the pre-specified headline family (\S6.5) with Holm--Bonferroni correction; no test in the family survives correction.

\begin{table}[t]
\centering
\caption{Pre-specified headline family (\S6.5) with Holm--Bonferroni correction. No test in the family survives correction.}
\label{tab:headline}
\small
\begin{tabular}{@{}lccccc@{}}
\toprule
Contrast & $\Delta$ accuracy [95\% CI] & $d_z$ & Perm.\ $p$ & Holm $p$ & Reject \\
\midrule
S0 vs.\ C1 & $+0.035$ [$-0.035$, $+0.110$] & 0.14 & 0.436 & 0.843 & no \\
C1 vs.\ C4 & $-0.050$ [$-0.160$, $+0.050$] & $-0.15$ & 0.422 & 0.843 & no \\
C2 vs.\ C1 & $+0.110$ [$+0.010$, $+0.210$] & 0.34 & 0.043 & 0.172 & no \\
$\mathrm{S0}'_{C2}$ vs.\ C2 & $-0.065$ [$-0.165$, $+0.025$] & $-0.20$ & 0.265 & 0.794 & no \\
\bottomrule
\end{tabular}
\end{table}

\textbf{RQ1 (falsification criterion \S6.6(1): triggered).} Both intermediates materially outperform the wide endpoint under the pre-stated bar --- C2 vs. C1: +0.110, CI [+0.010, +0.210], $d_z = 0.34$; C3 vs. C1: +0.110, CI [+0.005, +0.215], $d_z = 0.31$ (descriptive, outside the corrected family) --- but neither materially outperforms the fine endpoint: C2 vs. C4 is +0.060 with CI [$-0.005,$ +0.125] and C3 vs. C4 is +0.060 with CI [$-0.020,$ +0.135], both including zero, the former missing the bar by its lower bound. Per \S6.6(1), the intermediate-optimum hypothesis is therefore \textbf{unsupported for this workload}. We note the descriptive shape --- intermediates highest, endpoints lower --- is suggestive of the hypothesized non-monotonicity, and the C2/C4 interval only narrowly includes zero; at pilot scale this is an inconclusive reading, not evidence of absence (\S6.5).

\begin{figure}[t]
\centering
\includegraphics[width=0.92\linewidth]{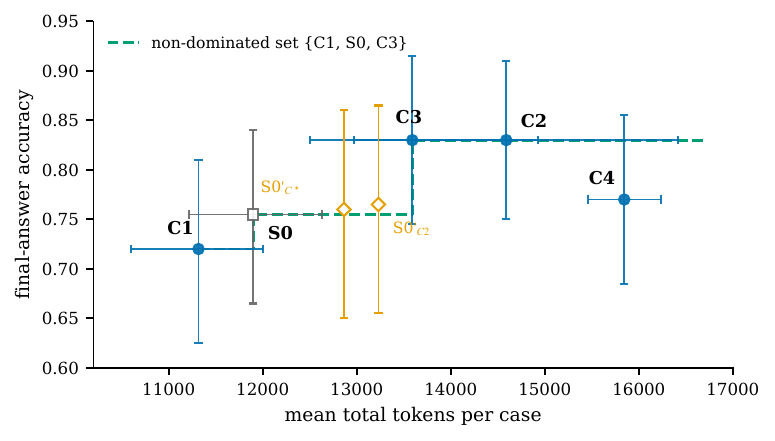}
\caption{Cost--quality plane: mean total tokens per case vs.\ final-answer accuracy, with 95\% CIs on both axes (Table~\ref{tab:main}). The dashed staircase traces the non-dominated set $\{$C1, S0, C3$\}$; open diamonds are the matched-token arms $\mathrm{S0}'_{C2}$ and $\mathrm{S0}'_{C^\star}$ (Table~\ref{tab:matched}). Each point aggregates $n = 40$ cases $\times$ five repeats (200 runs per condition).}
\label{fig:pareto}
\end{figure}

\textbf{RQ2 (falsification criterion \S6.6(2): not triggered).} S0 does not weakly Pareto-dominate C1--C4 on the (token-cost, accuracy) plane: C2 and C3 are more accurate at higher cost, and C1 is cheaper at lower accuracy. Descriptively, the non-dominated set is $\{$C1, S0, C3$\}$ (Figure~\ref{fig:pareto}): C3 weakly dominates C2 (equal accuracy, fewer tokens) and dominates C4 (higher accuracy, fewer tokens). 

The orchestrated multi-worker hypothesis survives its falsification test, at a measured price: the accuracy gain along the non-dominated set (S0 $\rightarrow$ C3, +0.075) costs +14\% tokens and +42\% median latency.

\subsection{Error localization}

Table~\ref{tab:errors} attributes each incorrect trace to its earliest failing subtask (\S6.1), over the 219 incorrect main-sweep runs.

\begin{table}[t]
\centering
\caption{Error localization: earliest failing subtask per incorrect main-sweep run (\S6.1), over the 219 incorrect runs.}
\label{tab:errors}
\small
\begin{tabular}{@{}lcccccc@{}}
\toprule
Condition & CLS & JUR & RAT & EXM & RCH & Total \\
\midrule
S0 & 0 & 30 & 0 & 0 & 19 & 49 \\
C1 & 0 & 23 & 0 & 0 & 33 & 56 \\
C2 & 0 & 0 & 15 & 0 & 19 & 34 \\
C3 & 0 & 7 & 10 & 0 & 17 & 34 \\
C4 & 0 & 6 & 1 & 0 & 39 & 46 \\
\bottomrule
\end{tabular}
\end{table}

CLS and EXM never appear as earliest failures in any condition. The wide-scope conditions fail earliest at jurisdiction determination (JUR: 30 of 49 for S0, 23 of 56 for C1); giving JUR a dedicated or co-resident-with-CLS worker eliminates JUR-earliest failures entirely under C2 (0) and nearly so under C3 and C4 (7 and 6). Fragmentation relocates rather than removes error: under C4, failures concentrate at the terminal synthesis step (RCH: 39 of 46), where the narrowest worker must reconcile all upstream records. This relocation pattern --- early-stage failures in wide scopes, synthesis failures in narrow scopes, with the intermediates minimizing totals (34 each) --- is the error-propagation texture behind the non-monotonic accuracy sequence, and we return to it in \S9.

\subsection{Matched-token comparison (RQ3)}

Table~\ref{tab:matched} reports the S0 family. Tuning used only the development split (\S6.3); evaluation-set usage is reported as observed.

\begin{table}[t]
\centering
\caption{Matched-token arms (\S6.3). Tuning used only the development split; evaluation-set usage is reported as observed.}
\label{tab:matched}
\small
\begin{tabular}{@{}lccccc@{}}
\toprule
Arm & Accuracy [95\% CI] & Tokens (eval) & Target & Deviation & In $\pm10\%$ band \\
\midrule
S0 & 0.755 [0.665, 0.840] & 11{,}899 & --- & --- & --- \\
$\mathrm{S0}'_{C2}$ & 0.765 [0.655, 0.865] & 13{,}230 & 14{,}590 & $-9.3\%$ & yes \\
$\mathrm{S0}'_{C^\star}$ & 0.760 [0.650, 0.860] & 12{,}864 & 13{,}591 & $-5.3\%$ & yes \\
\bottomrule
\end{tabular}
\end{table}

Three protocol disclosures. First, $C^\star$ resolved to \textbf{C3}: C2 and C3 tied exactly on mean accuracy (0.830), and the tie was broken by lowest mean token cost (13,591 < 14,590), a Pareto-consistent rule fixed only after the tie was observed and disclosed as such. Second, the tuning ladder is discrete, and both budget targets resolved to the same rung (extended role description, scratchpad instruction, three development-case exemplars); the two matched-token arms therefore ran an identical tuned prompt, and the post-selection arm functions as an \textbf{independent replication} of the confirmatory arm rather than a distinct budget point --- a replication it passes (0.765 vs. 0.760 on disjoint runs). Third, evaluation-set usage landed below both targets ($-9.3$\%, $-5.3$\%) but within the pre-stated $\pm$10\% band.

\textbf{RQ3 (falsification criterion \S6.6(3): triggered).} The confirmatory contrast $\mathrm{S0}'_{C2}$ vs. C2 is $-0.065$ with CI [$-0.165,$ +0.025], $d_z = -0.20$, permutation $p = 0.265$, Holm-adjusted $p = 0.794$: not significant, with an interval including zero. Per the pre-registered rule, we therefore report that \textbf{any C2 main-sweep advantage is consistent with a prompt-budget explanation rather than an agent-count explanation}. Per \S6.5, this is an inconclusive finding, not a null: the point estimate favors C2 --- the budget-matched single agent remains 6.5 points below it, at the small-effect threshold in magnitude --- and the added budget bought S0 essentially nothing (+0.010 and +0.005 accuracy for +11\% and +8\% tokens over plain S0). The data lean toward an agent-count effect, but at $n = 40$ the pilot cannot exclude the budget explanation, and the pre-registered criterion requires saying so.

Failure-injection results (RQ4) are reported in \S8.

\section{Localized Fallback under Injected Failures (RQ4)}

The three injection arms comprise 3,000 runs: for each mode --- forced worker timeout, hallucinated output, and shared-tool outage (\S6.4) --- every condition runs the full 40-case set with five repeats, paired by case identity and target subtask $\tau$ with the un-injected main sweep. Exactly one mode applies per run. The timeout arm reported here executed under the corrected forced-timeout dispatch; the original arm carried a harness defect, is archived with the artifact, and is not reported (\S10.2). RQ4 carries no falsification criterion: all quantities in this section are descriptive under the \S6.4 reporting specification, with paired per-case accuracy deltas and case-clustered 95\% bootstrap CIs (\S6.5).

\begin{figure}[t]
\centering
\includegraphics[width=\linewidth]{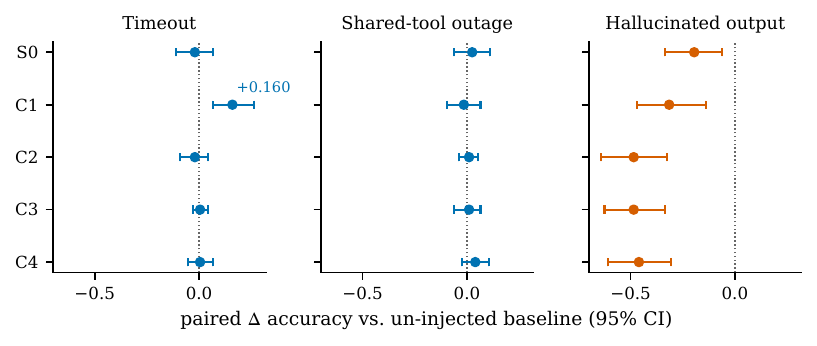}
\caption{Paired per-case accuracy deltas under injection, by condition, with case-clustered 95\% bootstrap CIs (Table~\ref{tab:injection}). Availability perturbations (left, center) straddle zero in every cell except the C1 timeout overshoot ($+0.160$); the content perturbation (right) is negative in every condition. Each cell aggregates $n = 40$ cases $\times$ five repeats (200 runs per (injection, condition) cell).}
\label{fig:injection}
\end{figure}

Table~\ref{tab:injection} reports the \S6.4 metrics per (injection, configuration) cell. \emph{Substitution success} --- the fraction of injected cases reaching a validated trace within the repair budget --- is computed among fired injections: all 40 cases for timeout and hallucination, the eight affected cases (20\%) for outage. $\Delta$ accuracy is the paired delta against the condition's own un-injected baseline (Table~\ref{tab:main}). Figure~\ref{fig:injection} plots the paired deltas.

\begin{table}[t]
\centering
\caption{Localized fallback under injection (\S6.4), per (injection, configuration) cell. $\Delta$ accuracy is the paired delta against the condition's own un-injected baseline (Table~\ref{tab:main}), with case-clustered 95\% bootstrap CIs.}
\label{tab:injection}
\footnotesize
\setlength{\tabcolsep}{4pt}
\begin{tabular}{@{}llccccc@{}}
\toprule
Injection & Cond. & Subst.\ success & All-case acc. & Val.-trace acc. & $\Delta$ acc.\ [95\% CI] & Token penalty \\
\midrule
\multirow{5}{*}{Timeout}
 & S0 & 1.000 & 0.735 & 0.735 & $-0.020$ [$-0.110$, $+0.065$] & $-1{,}442$ \\
 & C1 & 1.000 & 0.880 & 0.880 & $+0.160$ [$+0.065$, $+0.265$] & $+21{,}793$ \\
 & C2 & 0.965 & 0.810 & 0.839 & $-0.020$ [$-0.090$, $+0.045$] & $+4{,}611$ \\
 & C3 & 0.965 & 0.835 & 0.865 & $+0.005$ [$-0.030$, $+0.045$] & $+4{,}433$ \\
 & C4 & 0.995 & 0.775 & 0.779 & $+0.005$ [$-0.055$, $+0.065$] & $-87$ \\
\addlinespace
\multirow{5}{*}{Outage}
 & S0 & 1.000 & 0.780 & 0.780 & $+0.025$ [$-0.060$, $+0.110$] & $+1{,}138$ \\
 & C1 & 1.000 & 0.705 & 0.705 & $-0.015$ [$-0.095$, $+0.065$] & $+660$ \\
 & C2 & 1.000 & 0.840 & 0.853 & $+0.010$ [$-0.040$, $+0.055$] & $+363$ \\
 & C3 & 0.975 & 0.840 & 0.862 & $+0.010$ [$-0.060$, $+0.065$] & $+622$ \\
 & C4 & 1.000 & 0.810 & 0.814 & $+0.040$ [$-0.025$, $+0.105$] & $+623$ \\
\addlinespace
\multirow{5}{*}{Hallucination}
 & S0 & 1.000 & 0.560 & 0.560 & $-0.195$ [$-0.335$, $-0.060$] & $+3{,}290$ \\
 & C1 & 0.885 & 0.405 & 0.458 & $-0.315$ [$-0.470$, $-0.140$] & $+5{,}735$ \\
 & C2 & 0.900 & 0.345 & 0.383 & $-0.485$ [$-0.640$, $-0.325$] & $+2{,}152$ \\
 & C3 & 0.890 & 0.345 & 0.388 & $-0.485$ [$-0.625$, $-0.335$] & $+3{,}111$ \\
 & C4 & 0.925 & 0.310 & 0.335 & $-0.460$ [$-0.610$, $-0.305$] & $+2{,}009$ \\
\bottomrule
\end{tabular}
\end{table}

\subsection{Availability perturbations are absorbed}

Under shared-tool outage, every configuration recovers. Among the eight affected cases, 39 or 40 of the 40 injected runs per condition reach a validated trace (substitution success 0.975--1.000), all five paired deltas straddle zero, and token penalties are small (+363 to +1,138). S0 pays the largest penalty, consistent with whole-trace re-emission being the bluntest available retry instrument for a single transient tool call.

Forced timeout is likewise absorbed in four of five conditions. Substitution success is 0.965--1.000; the paired deltas for S0, C2, C3, and C4 all straddle zero. The token signature tracks the scope of the affected worker: the half-pipeline workers of C2 and C3 pay +4,433 to +4,611 tokens to regenerate their scope, the single-subtask worker of C4 pays essentially nothing ($-87$), and S0 lands slightly negative ($-1,442$) --- the forced initial dispatch bills no tokens, and the single repair regeneration replaces, rather than adds to, the baseline generation. A prediction recorded in the project log before the corrected arm was executed anticipated absorption in all five cells; four cells confirmed it, and the fifth exceeded it.

\subsection{The C1 timeout overshoot}

C1 under forced timeout reaches 0.880 --- the highest accuracy of any cell in the program, 0.160 above its own un-injected baseline, and the only availability cell whose interval excludes zero (CI [+0.065, +0.265]). The cost signature identifies what happened: +21,793 tokens --- a total of $\approx$2.9$\times$ the un-injected C1 cost --- i.e., the repair budget was consumed as roughly two additional full-pipeline attempts. After a forced restart, repair at full scope behaves as validation-filtered redraw of the entire pipeline, and the redraw more than recovers the discarded attempt. The effect scales with worker scope exactly as that account requires: full pipeline (C1, +0.160) $\gg$ half pipeline (C2/C3, $\approx$0) $\approx$ single subtask (C4, $\approx$0), and S0 --- whose normal repair unit is already the whole trace --- gains nothing ($-0.020$). Why validation-filtered redraw at wide scope outperforms the baseline's single-draft path, and does so only there, is taken up in \S9. We flag the finding's status explicitly: descriptive, at pilot scale, outside the corrected family.

\subsection{The content perturbation degrades every configuration and inverts the ordering}

Hallucinated output is the only mode in which every paired delta excludes zero, and all five are negative: S0 $-0.195$ [$-0.335,$ $-0.060$] is the mildest; C1 $-0.315$; C4 $-0.460$; C2 and C3, the main-sweep leaders, the worst at $-0.485.$ The main-sweep ordering inverts: S0 (0.560) is now the most accurate condition, C1 (0.405) second, and the fragmented configurations fall to 0.310--0.345.

Validation offers no content defense --- by construction (\S6.4), and the data confirm it emphatically. Injected runs almost always still reach a validated trace: substitution success is 0.885--0.925 for the orchestrated conditions and 1.000 for S0, and most of those validated traces are wrong (validated-trace accuracy 0.335--0.458 orchestrated). A schema-conforming, oracle-incorrect record passes $\mathcal{V}$ essentially unimpeded; as \S6.4 anticipated, recovery requires a downstream contradiction to surface, and where none does, the poison validates and ships.

The record substitution rate (\S6.4) exposes the sharpest tension in the program. The orchestrated conditions replace the injected record in 77.5--78.0\% of cases, S0 in only 58.0\% --- yet the orchestrated conditions end at 0.310--0.405 accuracy while S0 retains 0.560. Replacing the record is not correcting the determination: the replacement can be wrong again, and downstream workers that consumed the poisoned record before its replacement have already propagated its content across the hand-off. The record-based interfaces that give fragmented configurations their clean seams are, under content perturbation, equally clean propagation channels --- downstream workers treat validated upstream records as ground truth precisely because label isolation and schema validation license them to. Token penalties are moderate (+2,009 to +5,735), largest for C1, consistent with wide-scope redraws when contradictions do surface. We return to the propagation mechanism in \S9.

\subsection{Summary}

The three arms separate cleanly into a two-class taxonomy. \emph{Availability} perturbations --- forced timeout and transient shared-tool outage --- are absorbed by natural repair at every granularity tested, and at the widest orchestrated scope the restart is over-compensated into the program's best cell. The \emph{content} perturbation --- a single schema-conforming hallucinated record --- degrades every configuration, degrades the fragmented ones most, and inverts the main-sweep ordering. Localized fallback, as measured here, defends availability, not content.

\section{Discussion}

\textbf{Reading the curve through the chain-error model.} The sweep realizes the first expectation of \S5.4 in shape --- accuracy peaks at the intermediate configurations --- while failing the pre-registered material bar against the fine endpoint (\S7.1), so this pilot reads the non-monotonicity as directional rather than established. The error-relocation table (\S7.2) supplies the marginal-term evidence \S5.1 asked for: wide scopes concentrate earliest failures at jurisdiction determination (interference in $p_{\mathrm{JUR}}$ under broad visible state), the finest scope concentrates them at reverse-charge synthesis (fragmentation in $p_{\mathrm{RCH}}$ when the terminal worker must reconcile records it did not produce), and the intermediates minimize the total by placing their single boundary between the two error masses. Both signs that \S5.2 declined to fix materialized at once: wider was worse early, narrower was worse late. Consistent with boundary placement rather than worker count being the operative variable, C2 and C3 --- which share the downstream synthesis worker byte-for-byte and differ only in whether CLS and JUR are co-resident --- land on identical accuracy; the tie doubles as an internal-consistency check that the harness passes.

\textbf{Repair as resampling.} The timeout overshoot (\S8.2) gives the repair term $q_i^{(c)}$ of \S5.3 an unanticipated reading. A forced restart at full scope converts repair into validation-filtered redraw of the entire chain, and roughly two extra filtered drafts lifted C1 to 0.160 above its own baseline --- effective $\tilde{p}$ exceeding single-draft $p$ not through targeted correction but through selection over resamples. The effect requires wide scope (redraw at half- or single-subtask scope bought nothing) and is invisible to S0, whose whole-trace retry is already its normal mechanism ($Q^{(\mathrm{S0})}$). Validation-filtered resampling therefore looks like a reliability lever on the same axis the sweep varies --- at a measured $\approx$2.9$\times$ token cost --- and one separable from decomposition in a follow-up arm.

\textbf{Seams defend availability and propagate content.} \S5.3 scoped $q$ to validation-visible failures; the injection arms measure the complement directly. Availability faults were absorbed at every granularity (\S8.1), while the single schema-conforming content fault degraded every configuration and hit the fragmented ones hardest ($-0.460$ to $-0.485,$ against C1's $-0.315$). The mechanism is the hand-off itself: downstream workers treat validated upstream records as ground truth, so the record contract that gives orchestration its clean coordination is, under content perturbation, an equally clean propagation channel. Two cautions attach. Record replacement is not correction --- the orchestrated conditions replaced the poisoned record in $\approx$78\% of cases yet finished at 0.310--0.405 --- and the S0 row of that arm partly reflects the injection convention of \S6.4, under which S0's remaining fields predate the poison while orchestrated hand-offs consume it in-band; we therefore weight the within-orchestrated gradient, which the convention does not touch.

\textbf{What matched tokens did and did not buy.} The RQ3 verdict stands as pre-registered (\S7.3). What the arm adds mechanistically is that prompt budget was not S0's binding constraint: +8--11\% tokens of tuned role text, exemplars, and scratchpad moved accuracy by +0.005 to +0.010, while S0's modal earliest failure remained jurisdiction determination --- precisely the subtask the successful partitions isolate. The data lean toward an agent-count reading; at $n = 40$, the criterion rightly withholds it.

\textbf{A candidate heuristic, scoped to this workload.} For VAT determination on this rule set, model, and harness --- as measured, not as deployment guidance --- the pilot supports: place one boundary at the dependency-layer midpoint, separating preparation (CLS, JUR) from synthesis (RAT, EXM, RCH); prefer C3 over C2 on cost alone (equal accuracy, $\approx$7\% fewer tokens); do not fragment further, since the finest scope pays the synthesis penalty without an accuracy return; where transient availability faults dominate, wide scope with restart-on-stall was the strongest cell we measured; and at no granularity should schema validation be expected to stop content faults. Each clause is a hypothesis for the full study, not a rule. The pilot's sizing contribution is concrete: the effects worth confirming ran at $d_z \approx 0.3$, so a paired design with 0.8 power at $\alpha = .05$ requires on the order of 90 cases --- roughly twice this pilot --- before any clause graduates from heuristic to finding.

\section{Threats to Validity}
 
We organize threats by what each one limits: external validity, construct validity, statistical validity, and internal validity. The study is a bounded pilot: it is designed to isolate agent-scope granularity on one VAT-style activity surface, not to establish a general design rule for enterprise agents.
 
\subsection{External validity}
 
\textbf{Single workload.} The evaluation uses one bounded VAT activity surface (\S3). Findings should be read as a candidate heuristic for VAT-style determination workflows with oracle-labeled intermediate steps and a fixed rule set, not as a general claim about enterprise agents, tax automation, or legal compliance. The harness does not cover the full VAT rule space of any jurisdiction, special-sector schemes, real invoice parsing, supplier master-data defects, production posting, or statutory reporting (\S3.5). Cross-workload replication --- for example, single-jurisdiction sales tax, HS-code or tariff classification, claims adjudication, and KYC-style determinations --- is deferred to follow-on work.
 
\textbf{Synthetic cases and generator--oracle coupling.} The 40 evaluation cases are generated under the bounded rule set rather than sampled from production invoices (\S3.3). They exercise the specified activity surface but underrepresent real-world distribution shifts such as malformed inputs, ambiguous product descriptions, master-data errors, stale registration data, adversarial vendor behavior, and jurisdiction-specific edge cases. They also share an origin with the oracle: the same deterministic rule engine generates both cases and labels. This shared origin is necessary to obtain unambiguous step-level ground truth, but it can inflate measured accuracy if model outputs happen to align with rule-engine surface patterns rather than with the underlying VAT logic. Manual spot-checking (\S3.3) verifies oracle conformance within the harness; it does not establish production realism.
 
\textbf{Model homogeneity.} The main sweep uses one pinned base model family across C1--C4 and S0 (\S4.6). Heterogeneous-model variants are not run in this pilot. Conclusions about the granularity--performance relationship may shift under deployments where a stronger orchestrator model delegates to smaller or cheaper workers, or where different workers use different model families. We therefore report the result only for the homogeneous-model pilot and treat heterogeneous-model sweeps as outside the present scope.
 
\textbf{Model-version drift.} Model identifiers and decoding settings are pinned and released (\S6.7), which supports replication on the same versions but does not establish stability across providers, model generations, or capability tiers. The reported results should be interpreted as controlled comparisons among configurations under one pinned model family, not as absolute capability claims.
 
\textbf{Operating point of the model choice.} The pinned base model is a lower-cost tier, and that choice fixes a capability--difficulty operating point for the sweep rather than threatening internal validity: every condition faces the same model, and main-sweep accuracies (0.720--0.830) sit away from both floor and ceiling, so configuration differences remain observable. At other operating points the granularity curve could compress or reorder --- a stronger model may saturate the task toward ceiling and shrink the differences, while a weaker one may push conditions toward floor, where retry and fallback dominate behavior. The reported granularity effect is therefore an estimate at this operating point, and the difficulty band itself is among what a replication on another model tier would vary.
 
\subsection{Construct validity}
 
\textbf{Merge-policy dependence.} The C1--C4 partition follows one pre-specified dependency-layer merge policy (\S4.4). Alternative policies, including tool-locality clustering and semantic-similarity clustering, were considered but excluded from the primary sweep to avoid introducing a second experimental variable. The observed granularity pattern may therefore depend partly on the merge policy. Robustness across merge policies is deferred and reported as a limitation rather than implied by silence.
 
\textbf{Agent differentiation.} Section 4 defines agents as distinct by role, tool permissions, visible state, schema scope, and interaction protocol, not by prompt wording alone. This addresses the strongest version of the homogeneity critique \citep{xu2026oneflow}, but it does not prove that workers exhibit deep functional specialization. Under a single base model family, workers in C2--C4 may behave similarly despite different permissions and state slices. The pilot therefore measures the effect of architectural scope differences enforced by the harness, not specialization that would emerge from differentiated training, fine-tuning, or per-role model selection.
 
\textbf{Matched-token residual confounding.} The matched-token comparison (\S6.3) addresses the concern that multi-agent gains may reflect larger prompt budgets rather than agent count. It does not eliminate all structural differences. Even at matched token budgets, S0 and the orchestrated configurations differ in context segmentation, ledger state, validation localization, retry unit, and handoff structure. The matched-token result is therefore a sensitivity analysis for prompt-budget effects, not a complete causal decomposition of agent count, orchestration, and context structure.
 
\textbf{Matched-token band resolution.} The $\pm$10\% band check (\S6.3) is enforced on the development split with one repeat over eight cases, and measurement at that scale is coarse: the configuration ultimately selected measured roughly 15\% apart across the two independent tuning runs. Both budget targets consequently resolved to the same discrete tuning rung, so the two matched-token arms share one tuned prompt and the post-selection arm functions as an independent replication rather than a distinct budget point (\S7.3). Evaluation-set usage is reported as observed and landed within the band for both arms ($-9.3$\% and $-5.3$\%).
 
\textbf{S0 tuning.} S0 is tuned on an eight-case held-out development split (\S4.5), with the final prompt and tuning budget disclosed. A larger development set, different prompt-search method, or different whole-trace repair strategy could change S0 performance. We treat S0 as a strong, disclosed comparator rather than an exhaustively optimized single-agent upper bound. The development split also contains no exempt-outcome line, so S0's prompt is tuned without exposure to the exempt output shape that accounts for roughly 20\% of evaluation-set line determinations; the schema and output contract still specify that shape, but tuning-time coverage is absent.
 
\textbf{Failure-injection realism.} The fallback protocol (\S6.4) covers two isolated synthetic failures, timeout and hallucinated output, plus one correlated shared-tool outage with $m = 1$. Real deployments may experience more complex failures: provider-wide rate limiting, partial tool degradation, stale rate tables, refusal-policy interactions, cross-case context contamination, and upstream classification drift. The reported substitution-success and degradation metrics should therefore be read as localized-fallback measurements under controlled injections, not as a general fallback characterization across correlated failure modes. In addition, the injection unit follows each architecture's own structure (\S6.4): for S0 the subtask-$\tau$ slot is overwritten after the full trace is emitted, while orchestrated hand-offs consume the injected record in-band. The S0-versus-orchestrated contrast under hallucinated output therefore partly reflects this timing convention; contrasts among C1--C4, which share the convention, do not.
 
\textbf{Oracle scope.} The deterministic rule engine defines correctness only within the bounded VAT rule set (\S3.3). Oracle labels are not exposed to agents during generation or repair (\S6.2). The oracle can score emitted decisions against labels, but it cannot prove that a label-correct trace was reached through legally valid reasoning. A trace may match the oracle while containing superficial or coincidental reasoning. Step-level labels, rule-citation checks, and trace-consistency checks (\S3.4) reduce this risk but do not eliminate it. In addition, the rule set's exemption-over-reverse-charge precedence is never exercised by any evaluation or development case --- no exempt-category line occurs within an intra-community B2B transaction --- so the sweep cannot observe configuration behavior at that edge; a constructed probe confirms the oracle itself resolves it correctly.
 
\textbf{Harness defects discovered and repaired during execution.} Three defects in the experiment harness were discovered during the measured program --- each caught by the design's own instruments before analysis of the affected arm --- and each was repaired, with the arm re-run under the corrected harness. First, the activity-surface specification omitted the line-item operand from the reverse-charge worker's input state; the operand reached that worker only through a voluntary upstream echo that varied across partitions, deterministically failing amount-bearing cases under C2. The specification was amended under a closed-operand principle --- every field required by a subtask's contractual output belongs to that subtask's visible state (\S4.3) --- and the main sweep was re-run in full. Second, the forced-timeout injection initially skipped the worker dispatch entirely, so repair attempts ran against an empty conversation and the arm was uniformly fatal: it simulated ``task never delivered'' rather than the specified ``task delivered, call lost'' (\S6.4). The dispatch was corrected and the timeout arm re-run. Third, the matched-token arms initially executed with an untuned prompt identical to S0: the tuning step had been skipped and the harness fell back silently to the plain baseline. Both arms were re-run after tuning on the development split, and the silent fallback was replaced by a hard failure. All superseded attempts are archived in the released artifact alongside the repair history; only corrected-arm results are reported. The defects sit on the harness side of the construct: the oracle, dataset, and frozen hashes were unaffected throughout.
 
\subsection{Statistical validity}
 
\textbf{Pilot sample size.} Forty cases with five stochastic repeats per condition is a pilot-scale sample (\S6.5). We report paired bootstrap confidence intervals and effect sizes, including when intervals are wide. Non-significant comparisons in the headline family are interpreted as inconclusive rather than as evidence of no effect. We deliberately do not report p95 or p99 latency at this sample size; tail behavior requires a larger trial count. Per-scenario-family comparisons, drawing on ten cases each, are reported descriptively only.
 
\textbf{Multiplicity outside the headline family.} Holm--Bonferroni correction is applied only to the four pre-specified headline tests (\S6.5). Other comparisons, including post-selection sensitivity results, per-subtask error breakdowns, and best-versus-worst cost-quality contrasts, are descriptive. They should be interpreted as exploratory patterns rather than confirmatory hypothesis tests.
 
\textbf{Post-selection sensitivity.} The $\text{S0}'_{C^\star}$ comparison is selected after observing which orchestrated configuration has the highest mean final-answer accuracy (\S6.3). It is reported as a post-selection sensitivity analysis, not as a confirmatory test. The confirmatory matched-token comparison is $\text{S0}'_{C2}$ versus C2.
 
\textbf{Case-level aggregation.} Repeats are aggregated to case-level summaries before paired inference (\S6.5). This preserves the paired, case-clustered design and avoids treating stochastic repeats as independent cases. It also compresses repeat-level variance into a single case-level statistic. The raw repeat-level traces are released so that alternative analyses can be reproduced.
 
\subsection{Internal validity}
 
\textbf{Concurrency and provider effects.} The concurrency cap (\S4.6) prevents parallel-eligible workers from contending for provider-side rate-limit capacity. This avoids attributing provider queuing delays to agent granularity, but it also means the reported latencies do not characterize throughput-saturated deployments. Production settings with higher contention or provider-side throttling may shift the latency and cost frontier.
 
\textbf{Single orchestrator implementation.} All orchestrated conditions share one Magentic-One-style orchestrator (\S4.2). The S0--C1 contrast estimates the combined effect of adding the orchestrator, ledgers, validation loop, and subtask-level repair at one-worker scope. It does not isolate which component contributes which share of the effect. A different orchestrator design could change both the S0--C1 contrast and the C1--C4 trajectory.
 
\textbf{Repair-unit asymmetry.} C1--C4 repair at subtask granularity, while S0 repairs at whole-trace granularity (\S4.5). This asymmetry is intentional because it reflects the natural repair unit exposed by each architecture. It also means that retry-induced call count is not equal across conditions. We therefore report token usage, retry counts, and cost penalties explicitly rather than treating retries as hidden infrastructure.
 
\textbf{Validation visibility.} The validation checks in $\mathcal{V}$ detect schema errors, missing fields, missing citations, and some citation--decision inconsistencies. They do not detect all semantically wrong but schema-conforming outputs. This matters most in the hallucinated-output injection, where the injected record is intentionally validation-conforming (\S6.4). Such cases test silent semantic-error propagation rather than ordinary validation failure.
 
\textbf{Harness and prompt dependence.} Implementation choices may affect measured behavior: prompt wording, tool-call wrappers, timeout handling, retry ordering, and ledger serialization can all influence results. We release the harness, prompts, rule engine, and configuration files (\S6.7) so that these choices are inspectable, but releasing them does not neutralize their influence on the specific numbers reported here. Replication on the released artifact establishes that the pattern holds for this implementation; it does not establish that the pattern is implementation-independent.
 
The study therefore establishes a controlled pilot methodology and a bounded heuristic for VAT-style determination workflows. It does not establish a general design rule for LLM-agent decomposition, enterprise tax automation, or fault-tolerant agent systems.

\section{Conclusion}

We presented a pilot controlled sweep of agent-scope granularity on a fixed activity surface for synthetic EU-style VAT determination: a tuned no-orchestrator baseline and four orchestrated partitions of the same five subtasks, 40 oracle-labeled cases with five stochastic repeats, matched-token controls, and three failure-injection arms --- 4,400 valid runs evaluated against pre-registered falsification criteria. The intermediate configurations led on accuracy (0.830 for both) but did not meet the pre-stated material bar against the fine endpoint, so the intermediate-optimum hypothesis remains unsupported for this workload --- a directional reading at pilot scale, not evidence of absence; the single agent did not Pareto-dominate the orchestrated set; and the matched-token criterion fired, so any C2 main-sweep advantage is reported as consistent with a prompt-budget explanation, even though the point estimate favors C2. Under injection, availability faults were absorbed at every granularity --- with wide-scope restart over-recovering its own baseline by +0.160 --- while a single schema-conforming content fault degraded every configuration and inverted the main-sweep ordering. Together these results support one bounded heuristic --- a single partition boundary at the dependency-layer midpoint of a VAT-style workflow --- and one concrete sizing target for the full study, on the order of 90 cases; the oracle, dataset, harness, raw traces, and analysis pipeline are released to make both directly testable.

\paragraph{Artifact availability.}
The artifact --- oracle, dataset, harness, prompts, injection seeds, raw traces,
and analysis pipeline (\S6.7) --- is available at
\url{https://github.com/pedro-santos-eng/Right-sizing-LLM-agent-decomposition-in-VAT-determination} (archived at DOI \texttt{10.5281/zenodo.22083282}).

\bibliographystyle{plainnat}
\bibliography{references}

\newpage
\appendix
\section{S0 prompt and tuning disclosure}
\label{app:s0}
The S0 system prompt is assembled deterministically by \texttt{assemble\_s0\_prompt()} in the released artifact (\texttt{src/harness/s0.py}) from the shared static components of \S4 plus three S0-specific knobs; the text below is the assembled prompt used in the main sweep (SHA-256 \texttt{9d7ba74fa46456a4}).

{\scriptsize\begin{verbatim}
You are a VAT determination worker in a decomposed multi-agent system. You are responsible
for a fixed slice of the determination and nothing else. Work only from the state you are
given and the tools you may call. Do not restate table values from memory; obtain rates,
category kinds, and citation keys through the tools. Be deterministic and terse.

You solve the ENTIRE VAT determination for a case in one pass, with no orchestrator and no
per-subtask dispatch. You own all five subtasks: CLS, JUR, RAT, EXM, RCH.
You will receive the full case view (parties, transaction type, registration, and line
items).
Tools you may call: classification_reference, rate_table_lookup, rule_citation_retrieval,
vat_registration_check.

CLS — classification. Assign each line item to exactly one category from the closed
vocabulary. Call classification_reference to obtain the vocabulary and each category's
kind. Cite CLS.ASSIGNED.
JUR — jurisdiction / place of supply. From the supplier and customer countries, the
transaction type, and the customer's VAT-registration status, determine the jurisdiction
whose regime applies and the path taken (domestic / intra_community_b2b /
b2c_cross_border). You may call vat_registration_check for the registration input and
rule_citation_retrieval for the governing citation.
RAT — rate lookup. The rate band follows the classified category (bounded rule): reduced
band for RED_GOODS, RED_SERVICE; standard band for GEN_GOODS, GEN_SERVICE; no band for
EXEMPT_SUPPLY. For each line with a band, look up the rate for the jurisdiction and that
band with rate_table_lookup; never invent a rate. For an exempt-category line, no rate band
applies: record rate_band and rate as null and cite RATE.NA_EXEMPT.
EXM — exemption check. Using the exemption table provided in your state, decide whether
each line is exempt. An exempt line must cite the exemption rule; a non-exempt line is
taxed according to its band.
RCH — reverse-charge / liable-party synthesis. For each line resolve exactly one outcome
under the precedence EXEMPT > REVERSE_CHARGE > STANDARD_CHARGE, and set reverse_charge,
liable_party, vat_amount, and the non-charging reason accordingly. Cite the governing rule
obtained from rule_citation_retrieval.

CLS record: an object with keys subtask (== "CLS"), decision {category (one of:
"GEN_GOODS", "RED_GOODS", "GEN_SERVICE", "RED_SERVICE", "EXEMPT_SUPPLY")}, support (object
of the inputs you consumed), and rule_reference (exactly one of: "CLS.ASSIGNED").
JUR record: an object with keys subtask (== "JUR"), decision {jurisdiction (one of: "DE",
"FR", "IE", "ES"); jur_path (one of: "domestic", "intra_community_b2b",
"b2c_cross_border")}, support (object of the inputs you consumed), and rule_reference
(exactly one of: "JUR.B2C_CROSS_BORDER", "JUR.DOMESTIC", "JUR.INTRA_EU_B2B").
RAT record: an object with keys subtask (== "RAT"), decision {rate_band (string|null); rate
(number|null)}, support (object of the inputs you consumed), and rule_reference (exactly
one of: "RATE.NA_EXEMPT", "RATE.REDUCED", "RATE.STANDARD").
EXM record: an object with keys subtask (== "EXM"), decision {exempt (boolean)}, support
(object of the inputs you consumed), and rule_reference (exactly one of:
"EXM.EXEMPT_SUPPLY", "EXM.NONE").
RCH record: an object with keys subtask (== "RCH"), decision {outcome (one of:
"standard_charge", "reverse_charge", "exempt"); reverse_charge (boolean); liable_party (one
of: "supplier", "customer", "none"); vat_amount (number); non_charging_reason (one of:
"reverse_charge", "exempt", null)}, support (object of the inputs you consumed), and
rule_reference (exactly one of: "RC.B2B.INTRA_EU", "RC.B2C.SUPPLIER_CHARGES",
"RC.DOMESTIC.SUPPLIER_CHARGES", "RC.EXEMPT.NO_CHARGE").

Exemption table (reference R), authoritative:
EXEMPTION TABLE (bounded reference R). Only categories marked 'yes' are exempt.
category=GEN_GOODS exempt=no
category=RED_GOODS exempt=no
category=GEN_SERVICE exempt=no
category=RED_SERVICE exempt=no
category=EXEMPT_SUPPLY exempt=yes

Tool-use rules: cite only rule keys returned by rule_citation_retrieval; never guess table
values — call rate_table_lookup for rates and classification_reference for category kinds;
call vat_registration_check for the registration input where relevant.

Emit exactly one JSON object in a single fenced ```json block as the LAST content of your
final message, of the complete shape: {"case_id": <id>, "jur": <JUR record>, "lines":
[{"line_id": <id>, "cls": <record>, "rat": <record>, "exm": <record>, "rch": <record>},
...], "final": <aggregation block>}. Produce every record and the final block yourself.
"final": an object whose required keys are: currency (string); total_vat_amount (number);
lines (array). Optional: jurisdiction (one of: "DE", "FR", "IE", "ES"). Emit no other keys.
\end{verbatim}}

The matched-token arms tune three knob families on the eight-case development split (\texttt{dev\_001}--\texttt{dev\_008}): role-description expansion, worked exemplars, and an explicit scratchpad. The discrete tuning ladder resolved both budget targets ($\bar{B}_{C2}$ and $\bar{B}_{C^\star}$) to the same rung (L5), so both arms ran one tuned prompt (\S7.3); the ladder, the per-rung dev measurements, and the tuned prompt are in the artifact (\texttt{tune\_s0prime.py}). Per-worker prompt hashes for the orchestrated conditions:

\begin{center}\small\begin{tabular}{ll}\toprule Worker & SHA-256 (first 16) \\ \midrule
\texttt{C1:CLS{+}JUR{+}RAT{+}EXM{+}RCH} & \texttt{c57df552d55512b4} \\
\texttt{C2:CLS{+}JUR} & \texttt{29886613159ed705} \\
\texttt{C2:RAT{+}EXM{+}RCH} & \texttt{f01dc26662087cff} \\
\texttt{C3:CLS} & \texttt{e5757066e25a3950} \\
\texttt{C3:JUR} & \texttt{48b4aed6480e8fca} \\
\texttt{C3:RAT{+}EXM{+}RCH} & \texttt{f01dc26662087cff} \\
\texttt{C4:CLS} & \texttt{e5757066e25a3950} \\
\texttt{C4:EXM} & \texttt{1b2d9d2e6d74e957} \\
\texttt{C4:JUR} & \texttt{48b4aed6480e8fca} \\
\texttt{C4:RAT} & \texttt{aa5a3fbdb8566ce4} \\
\texttt{C4:RCH} & \texttt{8f563505f7016930} \\
\bottomrule\end{tabular}\end{center}

\section{Per-scenario-family descriptives}
\label{app:families}
Mean final-answer accuracy per scenario family in the main sweep (ten cases per family, five repeats aggregated to case level; descriptive only, no inference at $n{=}10$). Families derive deterministically from case inputs: multi-category invoices are \emph{mixed}; single-category cases split by supplier/customer country equality (\emph{domestic}) and transaction type (\emph{B2C} vs.\ \emph{intra-community B2B}). Row means reproduce Table~\ref{tab:main} exactly.

\begin{center}\small\begin{tabular}{lccccc}\toprule Family & S0 & C1 & C2 & C3 & C4 \\ \midrule
Domestic & 0.90 & 0.96 & 0.84 & 0.86 & 0.90 \\
Intra-community B2B & 0.82 & 0.70 & 0.88 & 0.92 & 0.62 \\
B2C rate differential & 0.68 & 0.62 & 0.84 & 0.82 & 0.86 \\
Mixed invoices & 0.62 & 0.60 & 0.76 & 0.72 & 0.70 \\
\bottomrule\end{tabular}\end{center}

The family texture matches the error localization of \S7.2: the wide scopes lead on \emph{domestic} (C1 0.96) but pay on \emph{intra-community B2B} (C1 0.70; C4 0.62), where jurisdiction and reverse-charge synthesis dominate; the intermediates are the only conditions above 0.70 in every family.

\section{Failure-injection configuration}
\label{app:injection}
All injection assignments are frozen in \texttt{data/injection\_plan.json} (generator L3-injection-v1, seed \texttt{20260801}); the harness consumes this file, so the assignments below are the executed ones. Per-case target subtask $\tau$ (shared across conditions and repeats):

{\scriptsize\begin{center}\begin{tabular}{llll}
\texttt{eval\_001}: RCH & \texttt{eval\_002}: RAT & \texttt{eval\_003}: RCH & \texttt{eval\_004}: JUR \\
\texttt{eval\_005}: CLS & \texttt{eval\_006}: JUR & \texttt{eval\_007}: RAT & \texttt{eval\_008}: JUR \\
\texttt{eval\_009}: RAT & \texttt{eval\_010}: CLS & \texttt{eval\_011}: RCH & \texttt{eval\_012}: CLS \\
\texttt{eval\_013}: JUR & \texttt{eval\_014}: RCH & \texttt{eval\_015}: EXM & \texttt{eval\_016}: JUR \\
\texttt{eval\_017}: RCH & \texttt{eval\_018}: RCH & \texttt{eval\_019}: RAT & \texttt{eval\_020}: RCH \\
\texttt{eval\_021}: RAT & \texttt{eval\_022}: JUR & \texttt{eval\_023}: JUR & \texttt{eval\_024}: JUR \\
\texttt{eval\_025}: CLS & \texttt{eval\_026}: RAT & \texttt{eval\_027}: JUR & \texttt{eval\_028}: RCH \\
\texttt{eval\_029}: JUR & \texttt{eval\_030}: CLS & \texttt{eval\_031}: EXM & \texttt{eval\_032}: RCH \\
\texttt{eval\_033}: EXM & \texttt{eval\_034}: CLS & \texttt{eval\_035}: JUR & \texttt{eval\_036}: CLS \\
\texttt{eval\_037}: EXM & \texttt{eval\_038}: RAT & \texttt{eval\_039}: CLS & \texttt{eval\_040}: RAT \\\end{tabular}\end{center}}

Shared-tool outage affects one seeded case per half-family block: \texttt{eval\_004}, \texttt{eval\_009}, \texttt{eval\_014}, \texttt{eval\_018}, \texttt{eval\_021}, \texttt{eval\_029}, \texttt{eval\_035}, \texttt{eval\_039}. Hallucinated records are constructed deterministically per case by the released generator: a schema-conforming record for $\tau$ whose decision contradicts the oracle label and whose rule citation is drawn from the closed citation set but does not support the decision; the full 40-record set ships in the plan file. Example (\texttt{eval\_001}):

{\scriptsize\begin{verbatim}
{
 "decision": {
  "liable_party": "customer",
  "non_charging_reason": "reverse_charge",
  "outcome": "reverse_charge",
  "reverse_charge": true,
  "vat_amount": 0.0
 },
 "rule_reference": "RC.B2B.INTRA_EU",
 "subtask": "RCH",
 "support": {
  "category": "GEN_SERVICE",
  "jur_path": "domestic"
 }
}
\end{verbatim}}

\section{Pinned model, decoding, and price sheet}
\label{app:pinned}
\begin{center}\small\begin{tabular}{ll}\toprule
Model (pinned) & \texttt{claude-haiku-4-5-20251001} \\
Temperature & 0.2 (top-$p$ at provider default; \S4) \\
Max tokens per call & 4096 \\
Per-call timeout & 120\,s \\
Per-subtask retry budget (C1--C4) & 3 \\
S0 whole-trace repair budget & 3 \\
Within-case concurrency cap & 2 \\
Price sheet (pinned 2026-08-02) & \$1.00 / \$5.00 per 1M input/output tokens \\
Oracle commit & \texttt{e2d2bdd22} \\
Dataset SHA-256 & \texttt{3dc683ec4186\ldots} \\
Generator seed & 42; cases 40 eval + 8 dev \\
\bottomrule\end{tabular}\end{center}

Dollar cost is derived from token counts at scoring time only; any price change during the study is a new pinned file version recorded in the project log.

\end{document}